\documentclass[pra]{revtex4}
\usepackage{eurosym}
\usepackage{amssymb}
\usepackage{amsmath}
\usepackage{graphicx}
\usepackage{dcolumn}
\usepackage[center]{subfigure}
\usepackage{color}

\begin{document}

\title{Two-dimensional solitons and quantum droplets supported by competing
self- and cross-interactions in spin-orbit-coupled condensates}
\author{Yongyao Li$^{1}$, Zhihuan Luo$^{2}$, Yan Liu$^{2}$, Zhaopin Chen$%
^{3} $, Chunqing Huang$^{1}$}
\email{Chunqinghuang@qq.com}
\author{Shenhe Fu$^{4}$, Haishu Tan$^{1}$}
\author{Boris A. Malomed$^{3,5,1}$}
\affiliation{$^{1}$School of Physics and Optoelectronic Engineering, Foshan University,
Foshan 528000, China \\
$^{2}$College of Electronic Engineering, South China Agricultural
University, Guangzhou 510642, China \\
$^{3}$ Department of Physical Electronics, School of Electrical Engineering,
Faculty of Engineering, Tel Aviv University, Tel Aviv 69978, Israel.\\
$^{4}$Department of Optoelectronic Engineering, Jinan University, Guangzhou
510632, China\\
$^{5}$Laboratory of Nonlinear-Optical Informatics, ITMO University, St.
Petersburg 197101, Russia}

\begin{abstract}
We study two-dimensional (2D) matter-wave solitons in spinor Bose-Einstein
condensates (BECs) under the action of the spin-orbit coupling (SOC) and
opposite signs of the self- and cross-interactions. Stable 2D two-component
solitons of the mixed-mode (MM) type are found if the cross-interaction
between the components is attractive, while the self-interaction is
repulsive in each component. Stable solitons of the semi-vortex type are
formed in the opposite case, under the action of competing self-attraction
and cross-repulsion. The solitons exist with the total norm taking values
below a collapse threshold. Further, in the case of the repulsive
self-interaction and inter-component attraction, stable 2D self-trapped
modes, which may be considered as quantum droplets (QDs), are created if the
beyond-mean-field Lee-Huang-Yang (LHY) terms are added to the self-repulsion
in the underlying system of coupled Gross-Pitaevskii equations. Stable QDs
of the MM type, of a large size with an anisotropic density profile, exist
with arbitrarily large values of the norm, as the LHY terms eliminate the
collapse. The effect of the SOC term on characteristics of the QDs is
systematically studied. We also address the existence and stability of QDs
in the case of SOC with mixed Rashba and Dresselhaus terms, which makes the
density profile of the QD more isotropic. Thus, QDs in the
spin-orbit-coupled binary BEC are for the first time studied in the present
work.\newline
\end{abstract}

\maketitle


\section{Introduction}

Producing stable bright solitons and solitary vortices in two- and
three-dimensional (2D and 3D) free space remains a challenging problem in
several areas of physics, such as nonlinear optics and Bose-Einstein
condensates (BECs). It is well known that the ubiquitous cubic
self-attractive nonlinearity readily predicts multidimensional solitons
which, however, are made completely unstable by the collapse. Diverse
methods have been proposed to stabilize 2D and 3D solitons, solitary
vortices, and more complex self-trapped states. In particular, stable 2D
optical solitons have been predicted and created in media with saturable
\cite{Segev1994}, quadratic \cite{Mihalache2006}, cubic-quintic \cite%
{Mihalache22006} and nonlocal nonlinearities \cite{Mihalache32006,Skupin},
which do not give rise to the collapse. Stable quasi-2D matter-wave solitons
were predicted in dipolar BECs with long-range interactions \cite%
{Pedri2005,Nath2008,Tikhonenkov2008,Yongyao2013,Tikhonenkov22008,Raghunandan2015,Jiasheng}%
.

It was predicted too that 2D and 3D solitons can be stabilized in spinor
(two-component) BECs with the help of the spin-orbit coupling (SOC) of the
Rashba type \cite%
{SVS1,SVS2,Bingjin2017,SVS3,SVSNJP,SVS4,Guihua2017,Yongchang,Yongping}.
There are two types of multidimensional solitons in this system, \textit{viz}%
., semi-vortices (SVs) and mixed modes (MMs). The SVs are characterized by
an exact ansatz with vorticities $\left( S_{+}=0,S_{-}=+1\right) $ [or its
flipped counterpart, with $\left( S_{+}=-1,S_{-}=+0\right) $] in its two
components, $\phi _{\pm }$ \cite{SVS1}, see Eq. (\ref{SVS}) below. MMs do
not admit a representation in the form of an exact ansatz, but they can be
constructed from an input which combines terms with vorticities $\left(
S_{+}=0;-1\right) $ and $\left( S_{-}=0;+1\right) $ in the two components
\cite{SVS1}, as per Eq. (\ref{mixed}), see below. Still more complex
self-trapped modes can be found as excited states built on top of the MMs,
starting from ansatz given below by Eq. (\ref{ESguess}). Excited states can
be also built on top of the SV, given by an exact ansatz with vorticities $%
\left( S_{+}=+1,S_{-}=+2\right) $ in the two components \cite{SVS1}, or its
flipped counterpart, with $\left( S_{+}=-2,S_{-}=-1\right) $. In the
previously studied systems, the excited states were found to be completely
unstable, unlike the fundamental MMs and SVs \cite{SVS1}.

Recently, we have found that SOC can also support stable 2D \textit{gap
solitons} in the free-space spinor BEC with dipole-dipole interactions
(possibly combined with contact interactions, which cannot create such
solitons by themselves) in the presence of the Zeeman splitting between the
components \cite{we2017}. In that case, the bandgap in the system's
spectrum, which is populated by the gap solitons, is a result of the
interplay of the SOC and Zeeman splitting. In nonlinear optics, SOC can be
emulated by dispersive coupling between parallel cores in twin planar
waveguides, which makes it possible to predict stable spatiotemporal
solitons (``light bullets") in the system \cite{SOCoptics}.

Thus, the two-component SOC\ systems offer a considerable potential for the
creation of stable self-trapped modes in 2D and even 3D \cite{Yongchang}
settings . Thus far, the studies of these systems dealt with the case when
both the self- and cross-component nonlinear interactions were attractive.
On the other hand, the relative sign of the interactions in the binary BECs
can be switched by means of the Feshbach resonance \cite{FR1,FR2}. In
particular, the opposite signs of the self- and cross-interactions make it
possible to predict one-dimensional solitons of the symbiotic type, which
may exist only in the two-component form \cite{symbio1,symbio2}.

One objective of this work is to consider 2D composite solitons in the
spinor BEC with SOC and competing self- and cross-interaction terms with
opposite signs. We find that, in the case of self-repulsion and
cross-attraction, stable MMs exist below a critical value of the total norm,
provided that the cross-attraction is stronger. Above the critical norm, the
MMs are destroyed by the collapse. In the opposite case of the
self-attraction and cross-repulsion, stable SVs exist with the total norm
smaller than the critical value corresponding to the Townes solitons \cite%
{Townes}.

Another possibility for the prediction of stable multidimensional
self-trapped states in the form \textquotedblleft quantum droplets" (QDs) is
offered by taking into account the Lee-Huang-Yang (LHY) corrections \cite%
{LHY} to the coupled mean-field Gross-Pitaevskii equations (GPEs) in binary
BEC. The self-repulsive LHY terms may compensate the contact attraction
between the components, which, otherwise, would destabilize the solitons due
to the possibility of the collapse \cite{Petrov2015,QDterm,Sadhan-nondip}.
Furthermore, the LHY terms were predicted \cite%
{Lima2011,Schutzhold2006,Saito2016,Oldziejewski2016,Bisset2016,Wachtler2016,Pastukhov2017,Cinti2017}
and experimentally demonstrated (in the ultracold gas of dysprosium atoms)
\cite{Chomaz2016,Kadau2016,Ferrier2016,Schmitt2016,Baillie2016,Edler2017} to
stabilize 2D and 3D self-trapped QDs supported by the long-range
dipole-dipole interactions. In particular, Ref. \cite{Schutzhold2006}
developed a general formalism for adding small corrections to the mean-field
equations. A very recent profoundly important experimental finding is that
3D QDs, with very low densities, can be also created via contact isotropic
interactions in the gas of $^{39}$K bosonic gas \cite{Leticia}.

The second objective of the present work is to consider the role of the LHY
terms in the 2D spinor condensate combining SOC with the contact
self-repulsion in each component and cross-attraction between them, or vice
versa (i.e., competing self- and cross- cubic terms). We find that MMs with
arbitrarily large norms exist as stable modes when the LHY correction terms
are present. The MMs in this case feature a large size and an anisotropic
density profile, which may be oriented in any direction in the system's 2D
plane, if the SOC is chosen the pure Rashba or pure Dresselhaus form (we
demonstrate that the interplay of mixed Rashba-Dresselhaus SOC with the LHY
terms also supports stable MM states, but the effective isotropy is broken
in that case). In the course of the analysis, we take into account the
effective renormalization of the strength of the LHY correction under the
action of SOC, that was demonstrated in Ref. \cite{Zhengwei2013}, in which,
however, the formation of QDs in spin-orbit-coupled BEC was not addressed.
In this connection, it relevant to mention that effects of another linear
interaction between the two components in the spinor BEC, \textit{viz}., the
Rabi coupling, was considered in Ref. \cite{Luca}, including the effect of
this interaction on the formation of the QDs under the action of the LHY
terms.

The paper is structured as follows. The model and some analysis of the
system are introduced in Sec. II. Stable MMs and their unstable excited
states (which may be weakly unstable) are studied by means of numerically
methods in Sec. III. Stable SVs, which are formed in case of the competing
self-attraction and cross-repulsion, are also considered in that section.
QDs of the MM type in the spin-orbit-coupled BEC are addressed in Sec. IV.
This is followed by the consideration of the interplay of the mixed
Rashba-Dresselhaus SOC with the LHY terms QDs of the MM type in the more
general SOC system, combining the Rashba and Dresselhaus couplings, are
considered in Sec. V, where it is found that QDs of the MM type persist and
may be stable even when the Rashba and Dresselhaus terms are equally mixed,
while the mean-field solitons (in the absence of the LHY corrections) do not
exist in the latter case. The paper is concluded by Sec. VI.

\section{The model}

In the scaled form, the 2D system of coupled GPEs for the spinor wave
function, $\boldsymbol{\phi }=(\phi _{+},\phi _{-})$, with competing self-
and cross-interaction local cubic terms (alias terms representing self- and
cross phase-modulation, i.e., SPM and XPM, respectively, in terms of
nonlinear optics \cite{Agrawal}), and SOC of the Rashba type, with strength $%
\lambda $, is written as 
\begin{eqnarray}
i\frac{\partial \phi _{+}}{\partial t} &=&-\frac{1}{2}\nabla ^{2}\phi
_{+}+(g|\phi _{+}|^{2}-\gamma |\phi _{-}|^{2})\phi _{+}+\lambda \hat{D}\phi
_{-},  \notag \\
i\frac{\partial \phi _{-}}{\partial t} &=&-\frac{1}{2}\nabla ^{2}\phi
_{-}+(g|\phi _{-}|^{2}-\gamma |\phi _{+}|^{2})\phi _{-}-\lambda \hat{D}%
^{\ast }\phi _{+},  \label{GPE}
\end{eqnarray}%
where the SOC operators may be written in the Cartesian or polar
coordinates:
\begin{eqnarray}
\hat{D} &=&\partial _{x}-i\partial _{y}\equiv e^{-i\theta }\left( \frac{%
\partial }{\partial r}-\frac{i}{r}\frac{\partial }{\partial \theta }\right) ,
\notag \\
\hat{D}^{\ast } &=&\partial _{x}+i\partial _{y}\equiv e^{i\theta }\left(
\frac{\partial }{\partial r}+\frac{i}{r}\frac{\partial }{\partial \theta }%
\right) ,  \label{D}
\end{eqnarray}%
while $g$ and $\gamma $ are effective 2D coupling constant \cite{QDterm}.
They are defined so that both are positive or negative, i.e., $g\gamma >0$.
For the time being, the LHY terms are not included. Note that the polar form
of the SOC operators, if substituted in Eq. (\ref{GPE}) demonstrates that,
if any anisotropic solution $\left\{ \phi _{+}(r,\theta ),\phi _{-}(r,\theta
_{-})\right\} $ is found, it generates a family of solutions \emph{rotated}
by an arbitrary angle, $\theta _{0}$, in the form of%
\begin{equation}
\left( \phi _{+},\phi _{-}\right) _{\mathrm{rotated}}=\phi _{+}\left(
r,\theta +\theta _{0}\right) ,~e^{-i\theta _{0}}\phi _{-}\left( r,\theta
+\theta _{0}\right) .  \label{rot}
\end{equation}

By means of further rescaling, one may set $\lambda =1$, and replace the
above-mentioned values of $g$ and $\gamma $ by either $\gamma =1$ (which we
adopt below in the case of $g,\gamma >0$), while $g>0$ is varied, or $g=-1$,
while $\gamma <0$ is varied (which is adopted below in the case of $g,\gamma
<0$). Note that $g=0$ (no SPM) may also correspond to the model of binary
Fermi gases \cite{Fermi}.

The energy of the system is 
\begin{gather}
E=\int \int \left\{ \frac{1}{2}\left( |\nabla \phi _{+}|^{2}+|\nabla \phi
_{-}|^{2}\right) +\frac{g}{2}\left( |\phi _{+}|^{4}+|\phi _{-}|^{4}\right)
-\gamma |\phi _{+}|^{2}|\phi _{-}|^{2}\right.  \notag \\
\left. +\lambda \left( \phi _{+}^{\ast }\hat{D}\phi _{-}-\phi _{-}^{\ast }%
\hat{D}^{\ast }\phi _{+}\right) \right\} dxdy.  \label{E}
\end{gather}%
In terms of the radial coordinate, $r=\sqrt{x^{2}+y^{2}}$, the asymptotic
form of the confined modes at $r\rightarrow \infty $ is looked for as $\phi
_{\pm }\sim \exp \left( -i\mu t-qr\right) $, where chemical potential $\mu $
is real, while radial wavenumber $q$ may be complex, with $\mathrm{Re}(q)>0$%
. The substitution of this in Eq. (\ref{GPE}) and linearization easily leads
to the expression for $q$ in terms of $\mu $:%
\begin{equation}
q=\pm i\lambda +\sqrt{2\mu +\lambda ^{2}}.  \label{q}
\end{equation}%
As it follows from here, the chemical potential satisfying the localization
condition, $\mathrm{Re}(q)>0$, takes values%
\begin{equation}
\mu <-\lambda ^{2}/2\equiv -1/2  \label{1/2}
\end{equation}%
(recall we have set $\lambda \equiv 1$). In particular, the limit case of
delocalized states, with $\mathrm{Re}(q)=0$, corresponds to $\mu =-1/2$.

In the case of $g,\gamma >0$, which corresponds to the competition between
the self-repulsion and cross-attraction (this sign combination may give rise
to symbiotic solitons in the 1D geometry without SOC \cite{symbio1,symbio2}%
), the system does not support SVs, but it readily creates MMs, with equal
norms in the two components,
\begin{equation}
N_{+}=N_{-}\equiv \int \int |\phi _{\pm }\left( x,y\right) |^{2}dxdy.
\label{N+-}
\end{equation}%
It is relevant to mention that, in terms of the total norm, $N\equiv
N_{+}+N_{-}$, the threshold for the onset of the critical 2D collapse in the
framework of Eqs. (\ref{GPE}) is%
\begin{equation}
N_{\mathrm{thr}}(g)=2\left( \gamma -g\right) ^{-1}N_{\mathrm{Townes}}\approx
11.70\left( \gamma -g\right) ^{-1},  \label{th}
\end{equation}%
where $N_{\mathrm{Townes}}\approx 5.85$ is the norm of single-component
Townes soliton \cite{Townes}-\cite{Fibich}. This value, as the boundary
between stable and collapsing MM states, is completely corroborated by
numerical results. 

The initial ansatz for the fundamental MM is adopted, in polar coordinates $%
\left( r,\theta \right) $, as in Ref. \cite{SVS1}, i.e.,%
\begin{eqnarray}
\phi _{+}(t &=&0)=A_{1}\exp \left( -\alpha _{1}r^{2}\right) -A_{2}r\exp
\left( -i\theta -\alpha _{2}r^{2}\right) ,  \notag \\
\phi _{-}(t &=&0)=A_{1}\exp \left( -\alpha _{1}r^{2}\right) +A_{2}r\exp
\left( i\theta -\alpha _{2}r^{2}\right) ,  \label{mixed}
\end{eqnarray}%
with $\alpha _{1,2}>0$ and real amplitudes $A_{1,2}$. 
Further, following Ref. \cite{SVS1}, it is possible to construct excited
states built on top of the MM, starting from ansatz

\begin{eqnarray}
\phi _{+}(t &=&0)=A_{1}r\exp \left( i\theta -\alpha _{1}r^{2}\right)
-A_{2}r^{2}\exp \left( -2i\theta -\alpha _{2}r^{2}\right) ,  \notag \\
\phi _{-}(t &=&0)=A_{1}r\exp \left( -i\theta -\alpha _{1}r^{2}\right)
+A_{2}r^{2}\exp \left( 2i\theta -\alpha _{2}r^{2}\right) .  \label{ESguess}
\end{eqnarray}%
These \textit{ans\"{a}tze} are used below to obtain numerically exact
solutions for fundamental and excited MMs.


Unlike MMs, self-trapped states in the form of SVs correspond to an exact
ansatz, which is compatible with Eq. (\ref{GPE}) \cite{SVS1}:%
\begin{equation}
\phi _{+}=e^{-i\mu t}f_{+}(r),~\phi _{-}=e^{-i\mu t+i\theta }rf_{-}(r),
\label{exact}
\end{equation}%
with chemical potential $\mu <-\lambda ^{2}/2$ and real amplitude functions $%
f_{\pm }(r)$, with nonzero values at $r=0$, which decay exponentially, as $%
\exp \left( -\sqrt{-2\mu -\lambda ^{2}}r\right) $ at $r\rightarrow \infty $,
see Eq. (\ref{q}). Below, we use the input in the form of
\begin{equation}
\phi _{+}(t=0)=A_{1}\exp (-\alpha _{1}r^{2}),\quad \phi _{-}(t=0)=A_{2}r\exp
(i\theta -\alpha _{2}r^{2}),  \label{SVS}
\end{equation}%
with real amplitudes $A_{1,2}$ and $\alpha _{1,2}>0$, to construct
numerically exact SV solutions of Eq. (\ref{GPE}).

It is relevant to stress that the norm of the wave function, $N$, defined
above, is proportional to the number of atoms in the condensate, but is not
identical to it. To estimate the actual number of atoms, we notice that,
according to Ref. \cite{SVS3}, the comparison of scaled 2D equations (\ref%
{GPE}) with the underlying system of 3D GPEs written in physical units shows
that the unit length in the scaled equations equations corresponds to the
physical length about 1 $\mathrm{\mu }$m. Further, assuming typical values
of the transverse confinement length $\simeq $ 3 $\mathrm{\mu }$m and
scattering length $\sim -0.1$ nm for the interatomic attraction, we conclude
that $N=1$ in the present notation is tantamount to $\simeq 3000$ atoms.

\section{Numerical results for the mean-field system (without the LHY terms)}

\subsection{Stable fundamental mixed modes}

\begin{figure}[t]
\subfigure[]{\includegraphics[width=0.3\columnwidth]{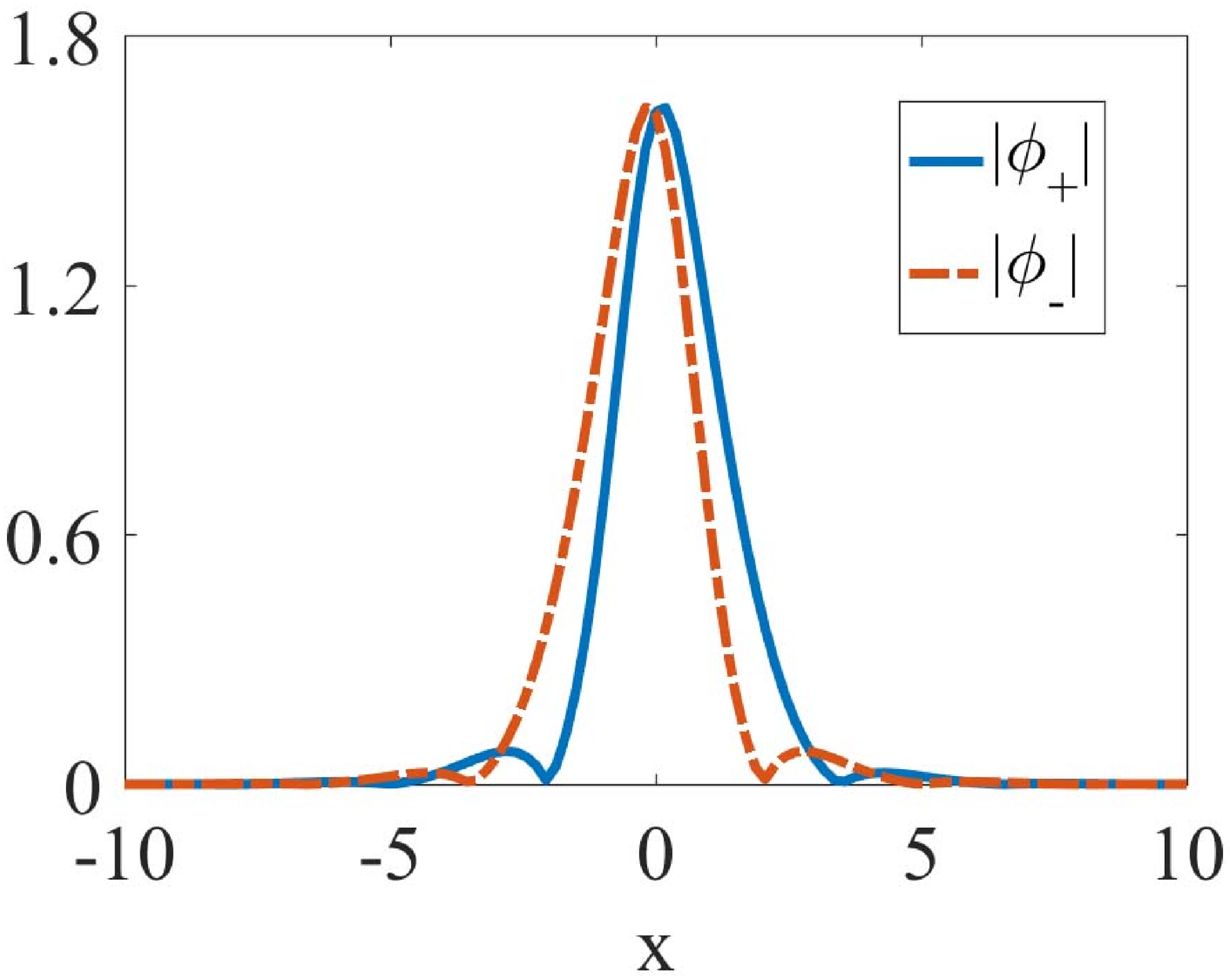}} %
\subfigure[]{\includegraphics[width=0.3\columnwidth]{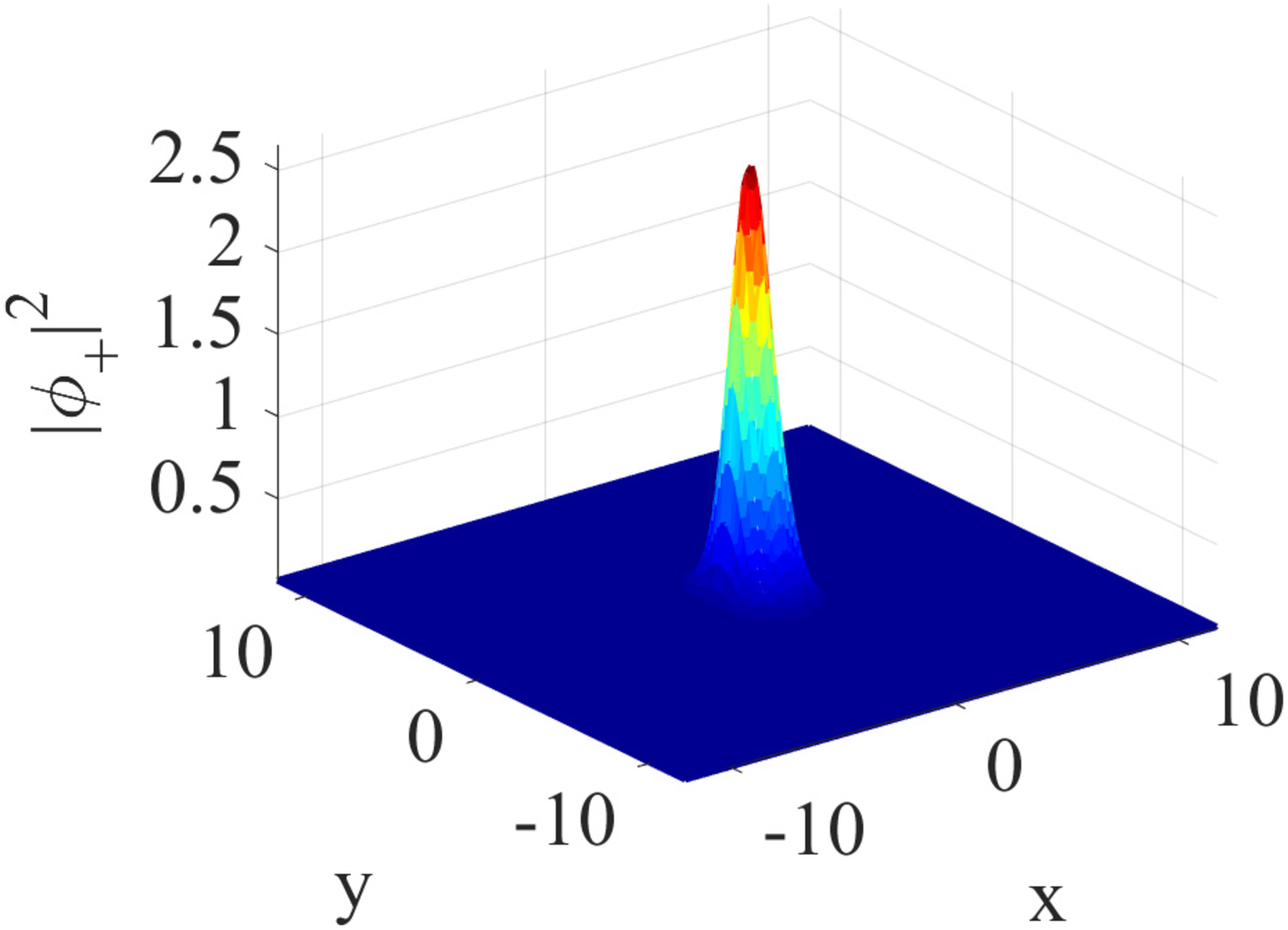}} %
\subfigure[]{\includegraphics[width=0.3\columnwidth]{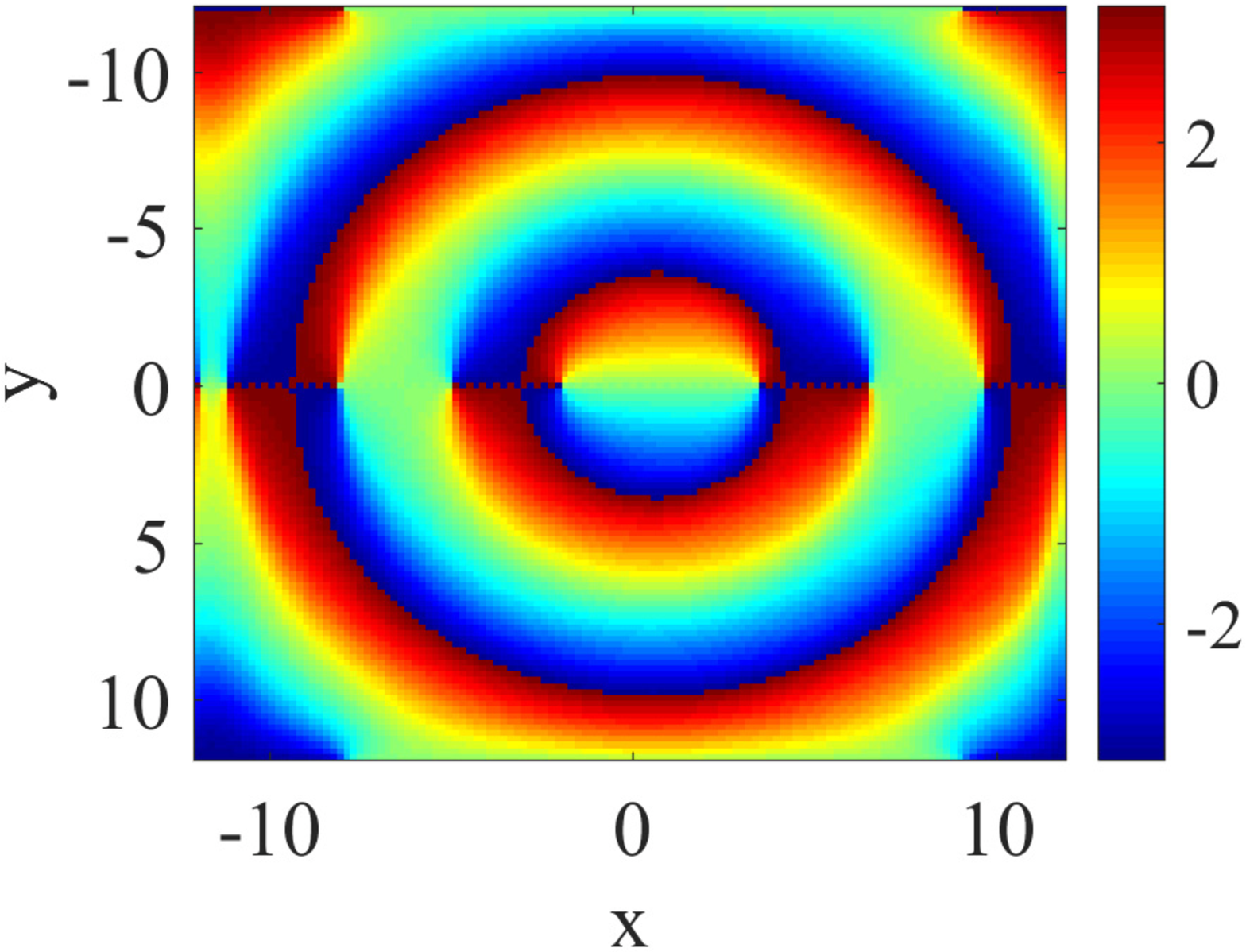}}
\caption{An example of a stable MM (mixed mode), obtained for $g=0.5$ and $%
\protect\lambda =\protect\gamma =1$ (these values of the SOC and XPM
coefficients are fixed by rescaling in all figures which report results for
MMs). (a) One-dimensional cross sections of the two components, $|\protect%
\phi _{+}|$ and $|\protect\phi _{-}|$, of the two-dimensional MM. The total
norm of the solution is $N=20$. (b,c) The density and phase structure of
component $\protect\phi _{+}$ of the same solution.}
\label{HS}
\end{figure}

By means of the imaginary-time method \cite{ITP1,ITP2}, MM solutions to Eq. (%
\ref{GPE}) can be generated from initial guess (\ref{mixed}). A typical
example of the so produced MM with $(N,g,\gamma )=(20,0.5,1)$ is displayed
in Fig. \ref{HS}. Note a small separation between maxima of the two
components in Fig. \ref{HS}(a), which is a characteristic feature of MMs
\cite{SVS1}.

\begin{figure}[t]
\subfigure[]{\includegraphics[width=0.3\columnwidth]{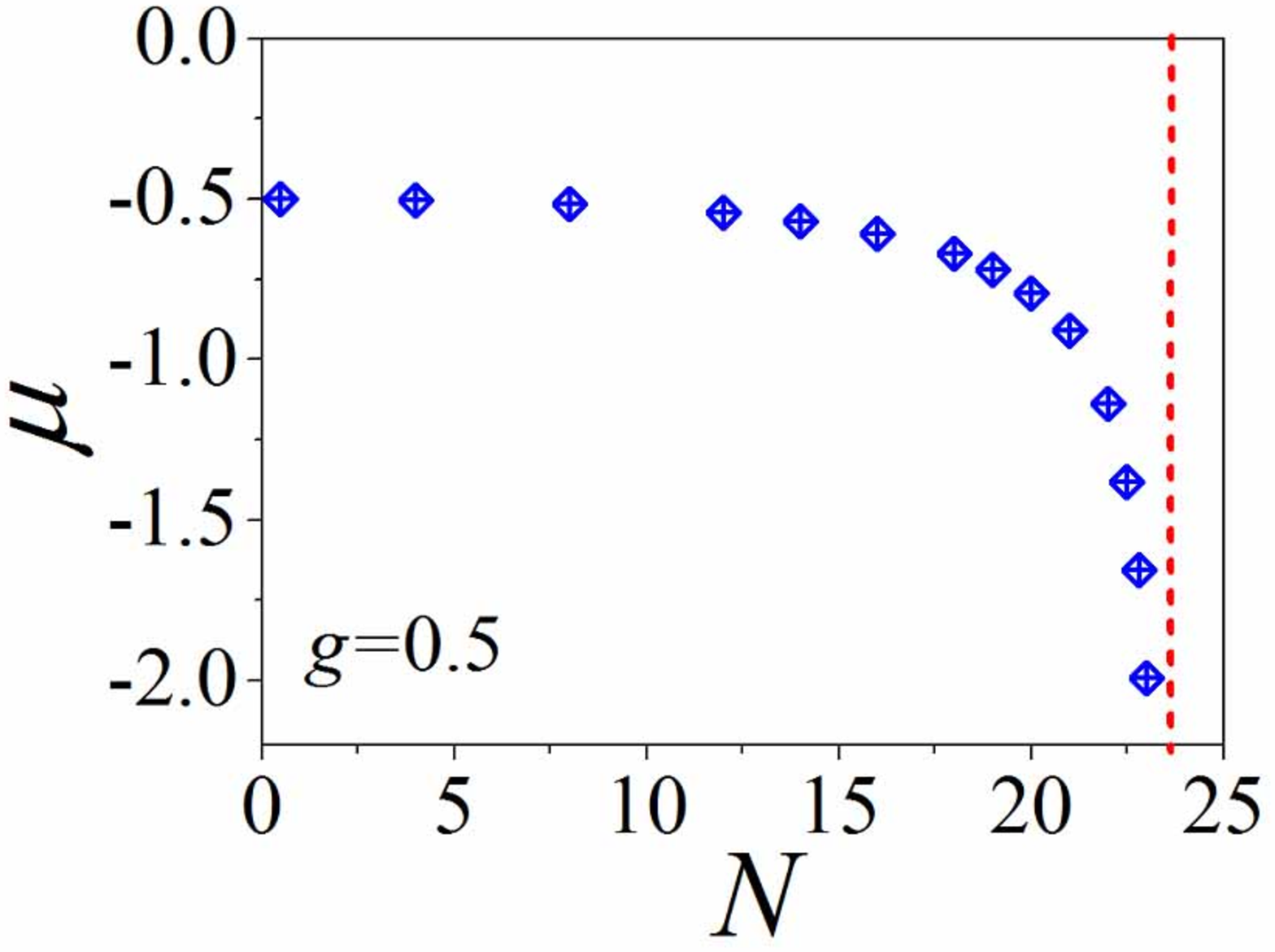}} %
\subfigure[]{\includegraphics[width=0.3\columnwidth]{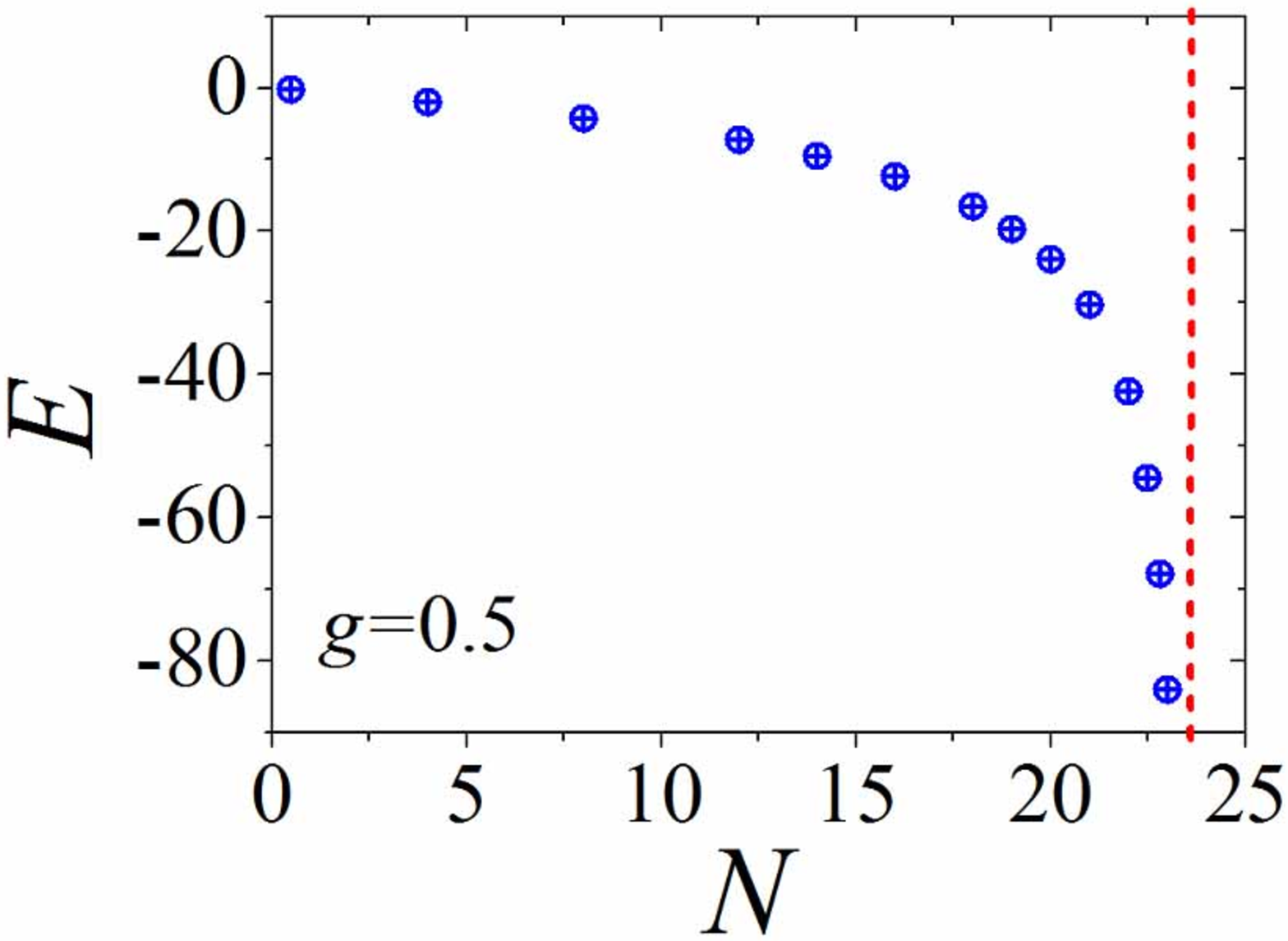}}
\caption{The chemical potential $\protect\mu $ (a) and energy $E$ (b) of the
MM mode vs. $N$ at $g=0.5$. The red dashed line in (a) is the existence
boundary, $N=N_{\mathrm{thr}}(g=0.5)\approx 23.4$, as given by Eq. (\protect
\ref{th}).}
\label{Numerics}
\end{figure}

The numerical solution produces a family of stable MMs in the interval of $%
0\leq g<\gamma \equiv 1$ [recall $\gamma \equiv 1$ is imposed in Eq. (\ref%
{GPE}) by rescaling]. In precise agreement with Eq. (\ref{th}), for fixed $%
g<1$ stable MMs are found at $N<N_{\mathrm{thr}}(g)$, and input in the form
of Eq. (\ref{mixed}) suffers collapse at $N>N_{\mathrm{thr}}(g)$. The
chemical potential, $\mu $, and energy (\ref{E}) of the MMs are displayed
versus $N$ in Figs. \ref{Numerics}(a) and (b), respectively. In the limit of
$N\rightarrow 0$, the MM turns into an infinitely broad state with an
infinitely small amplitude and the above-mentioned limit value of the
chemical potential, $\mu =-1/2$. Note that Fig. \ref{Numerics}(a)
demonstrates that the MM family, which is supported by the dominant
attractive XPM, satisfies the Vakhitov-Kolokolov (VK) criterion, $d\nu /dN<0$%
, that, in turn, is a well-known necessary condition for the stability of
the self-trapped modes \cite{VaKo}-\cite{Fibich}. \

\begin{figure}[t]
\subfigure[]{\includegraphics[width=0.3\columnwidth]{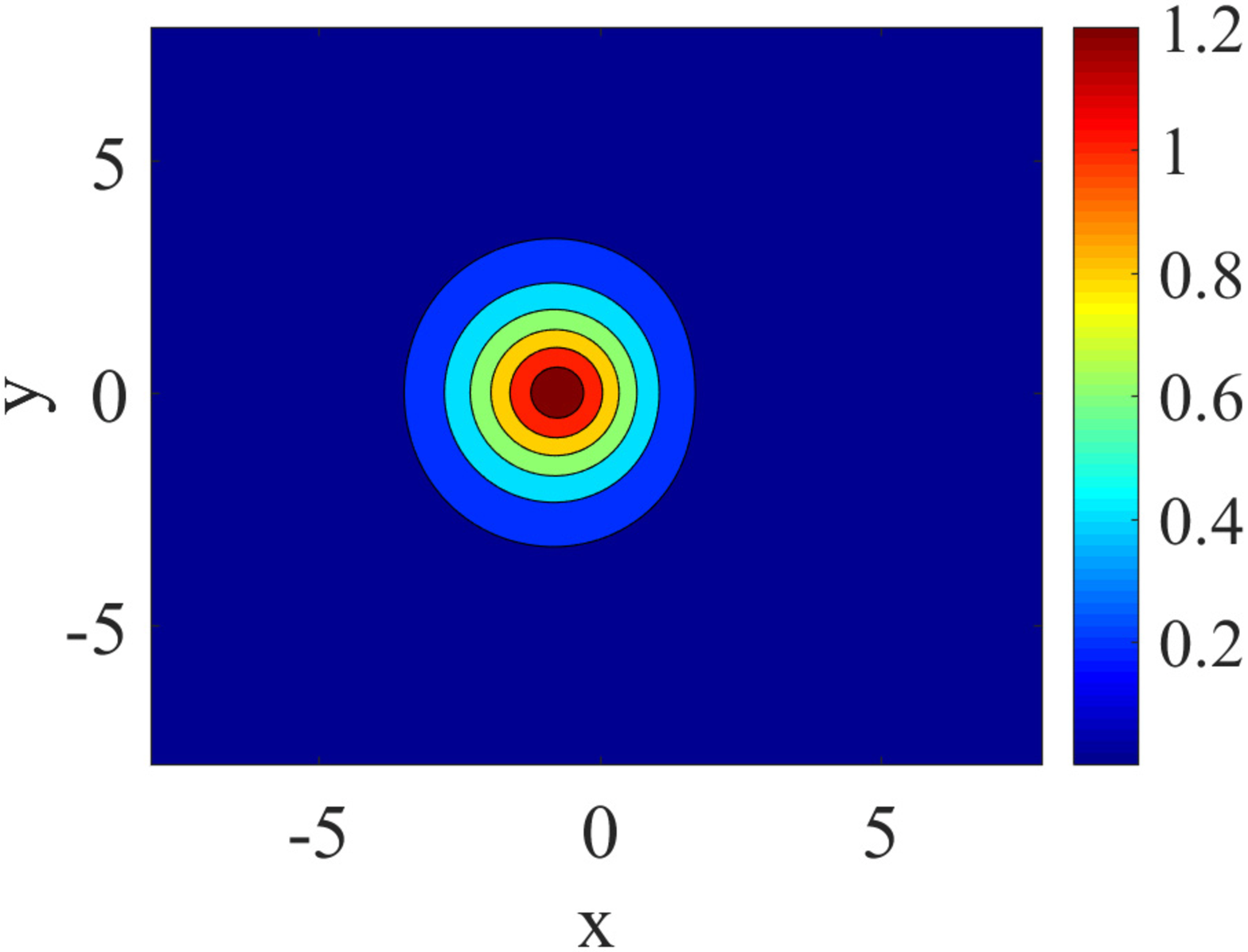}} %
\subfigure[]{\includegraphics[width=0.3\columnwidth]{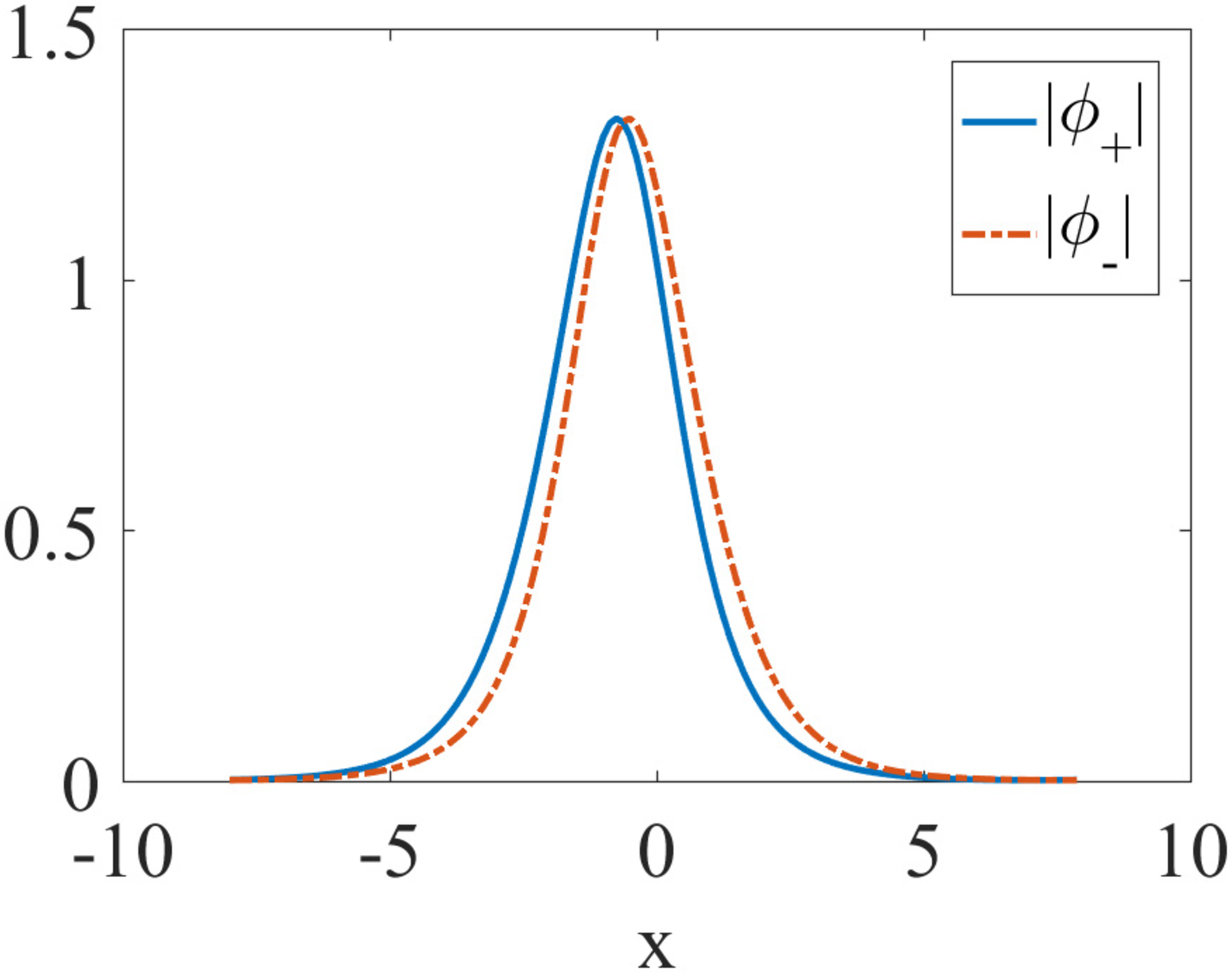}} %
\subfigure[]{\includegraphics[width=0.3\columnwidth]{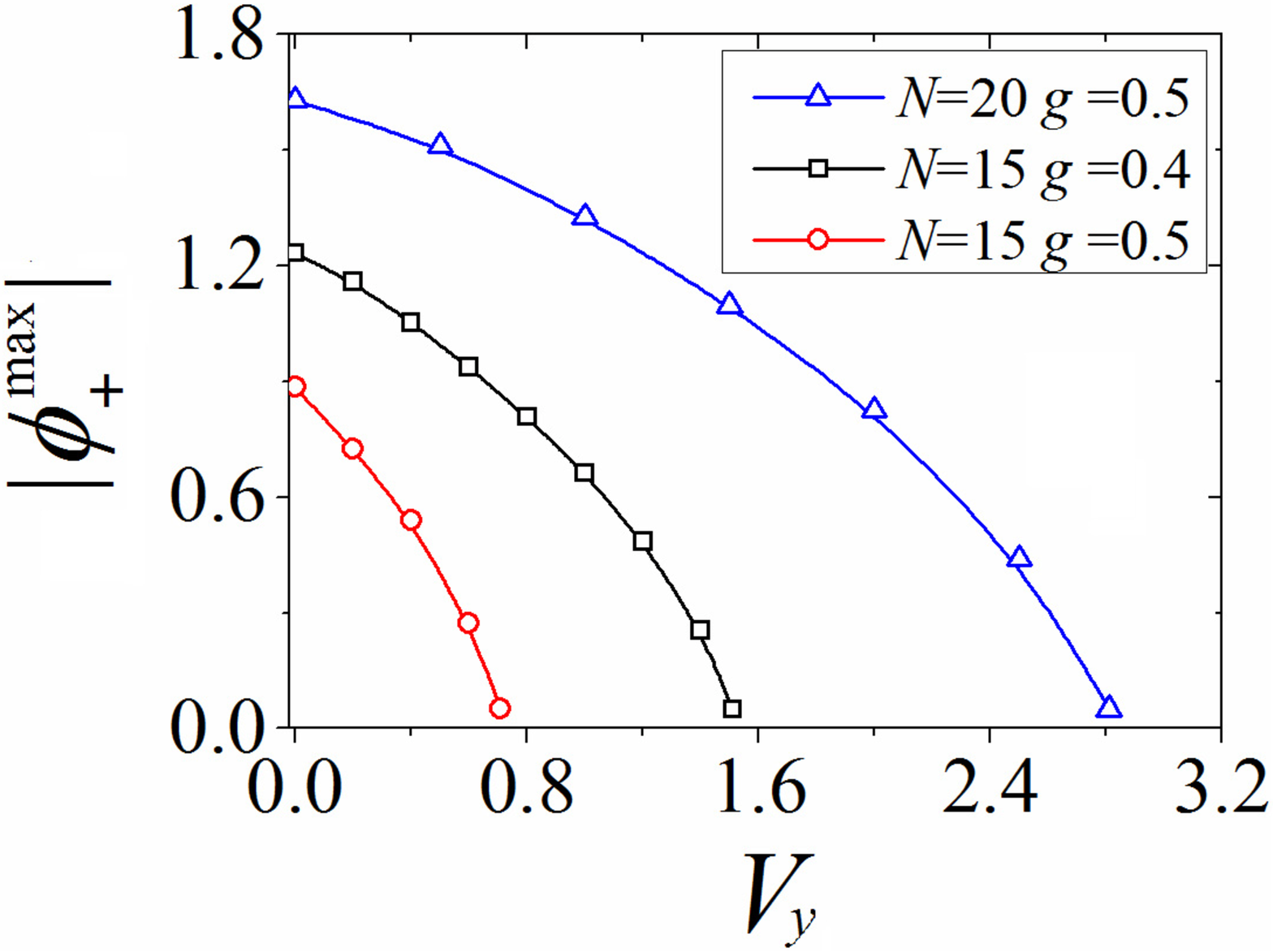}}
\caption{The 2D shape of $|\protect\phi _{+}|$ (a) and cross-section $|%
\protect\phi _{\pm }(x,0)|$ (b) of a stable moving MM, with $%
(N,g,V_{y})=(20,0.5,1)$. (c) The amplitude of $|\protect\phi _{+}|$ as a
function of velocity $V_{y}$ for different values of $N$ and $g$.}
\label{Vy}
\end{figure}

Mobility of the MMs is a nontrivial issue, as Eq. (\ref{GPE}) is not
Galilean invariant. Following the pattern of Ref. \cite{SVS1}, the mobility
was studied by rewriting Eq. (\ref{GPE}) in the reference frame moving with
velocity $\mathbf{V}=(V_{x},V_{y})$: 
\begin{eqnarray}
i\frac{\partial \phi _{+}}{\partial t} &=&-\frac{1}{2}\nabla ^{2}\phi _{+}+i(%
\mathbf{V}\cdot \nabla )\phi _{+}+(g|\phi _{+}|^{2}-\gamma |\phi
_{-}|^{2})\phi _{+}+\lambda \hat{D}\phi _{-},  \notag \\
i\frac{\partial \phi _{-}}{\partial t} &=&-\frac{1}{2}\nabla ^{2}\phi _{-}+i(%
\mathbf{V}\cdot \nabla )\phi _{-}+(g|\phi _{-}|^{2}-\gamma |\phi
_{+}|^{2})\phi _{-}-\lambda \hat{D}^{\ast }\phi _{+},  \label{movingGPE}
\end{eqnarray}%
Here we only consider the case of $V_{x}=0$, because the imaginary-time
method could produce stationary solution of Eq. (\ref{movingGPE}) solely in
this case, cf. Ref. \cite{SVS1}. Figures \ref{Vy}(a,b) present a typical
stable solution of Eq. (\ref{movingGPE}) for $N=20$ and $g=0.5$. Note weakly
anisotropic deformation of the 2D profile caused by the steady motion in the
$y$ direction.

The numerical solution demonstrates that, for fixing $N$ and $g$, the
amplitude of the moving MM, $|\phi _{\pm }^{\max }|$, decays (while it width
increases) with the increase of $V_{y}$, as shown in Fig. \ref{Vy}(c). In
this panel, the amplitude vanishes at $V_{y}^{\max }\approx 0.72$, $1.52$,
and $2.83$, for $(P,g)=(15,0.5)$, $(15,0.4)$, and $(20,0.5)$, respectively,
and the MMs do not exists at $V_{y}$ exceeding these limit values. Thus, the
MM's mobility range expands with the increase of the norm and decrease of $g$%
, up to $g=0$. The asymmetry of the moving solitons remains weak even when $%
V_{y}$ is close to the limit value. It is relevant to mention that the
mobility remains restricted to final values of $V_{y}$ also in the system
with the self-attractive SPM, i.e., $g<0$ \cite{SVS1}. In agreement with a
qualitative analysis of Eq. (\ref{movingGPE}), moving solitons with $V_{y}<0$
are mirror images, with respect to the $y$-axis, of their counterparts with $%
V_{y=0}>0$. Accordingly, the dependence of their amplitude on $\left\vert
V_{y}\right\vert $ is the same as shown in Fig. \ref{Vy}(c).

\subsection{Unstable excited mixed-mode states}

\begin{figure}[t]
{\includegraphics[width=0.9\columnwidth]{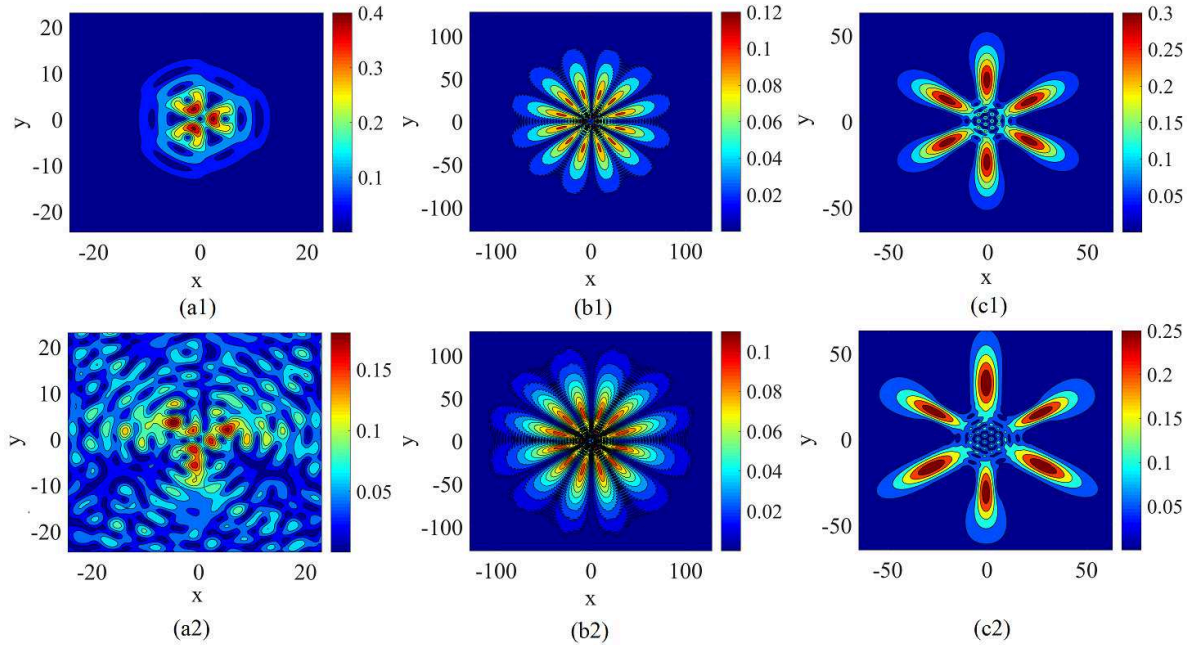}}
\caption{(a1) The stationary shape of an excited MM with $(g,N)=(0.5,20)$.
(a2) The result of the unstable evolution of this mode at $t=1000$. Panels
(b1,b2) and (c1,c2) show the same for $(g,N)=(0.9,100)$ and $(0.9,200)$,
respectively. For better visibility, profiles of the absolute value of the
field, $|\protect\phi _{+}(x,y;t)|$, are displayed here, instead of its
density, $|\protect\phi _{+}(x,y;t)|^{2}$.}
\label{ESNumerics}
\end{figure}

By means of the squared-operator method \cite{SOM}, excited MMs, initiated
by ansatz (\ref{ESguess}), have been found too. Numerical simulations
demonstrate that all such excited states are unstable. Typical examples of
their shape and evolution, shown in Fig. \ref{ESNumerics}, demonstrate that,
with the increase of $g$, the mode's shape becomes sharper and essentially
less unstable, transforming from a ``trefoil" in (a) to a 12-petal
``camomile" in (b) to a 6-vane ``windmill" in (c). In particular, the latter
pattern seems nearly stable.

\subsection{Stable semi-vortices}

\begin{figure}[t]
\subfigure[]{\includegraphics[width=0.3\columnwidth]{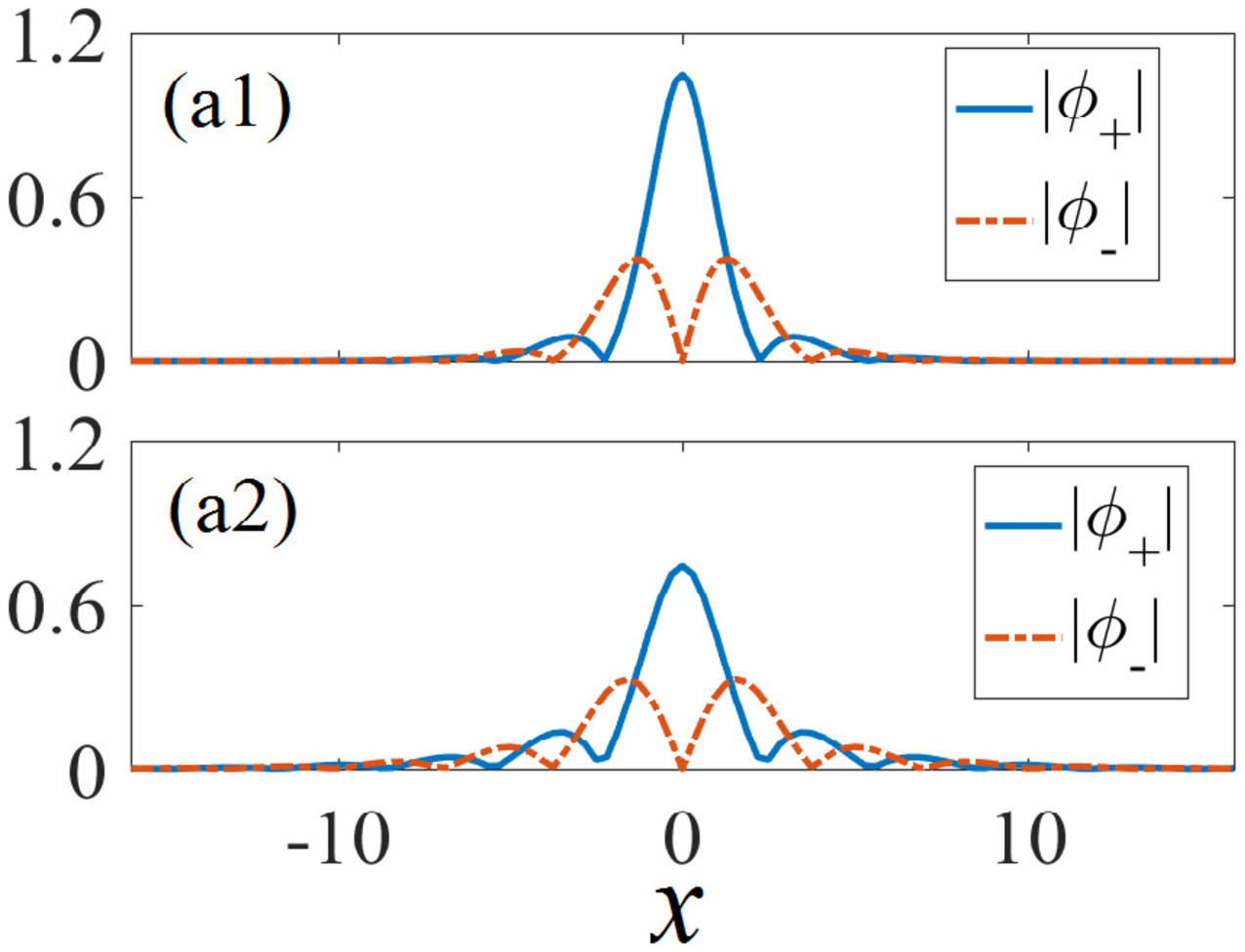}} %
\subfigure[]{\includegraphics[width=0.3\columnwidth]{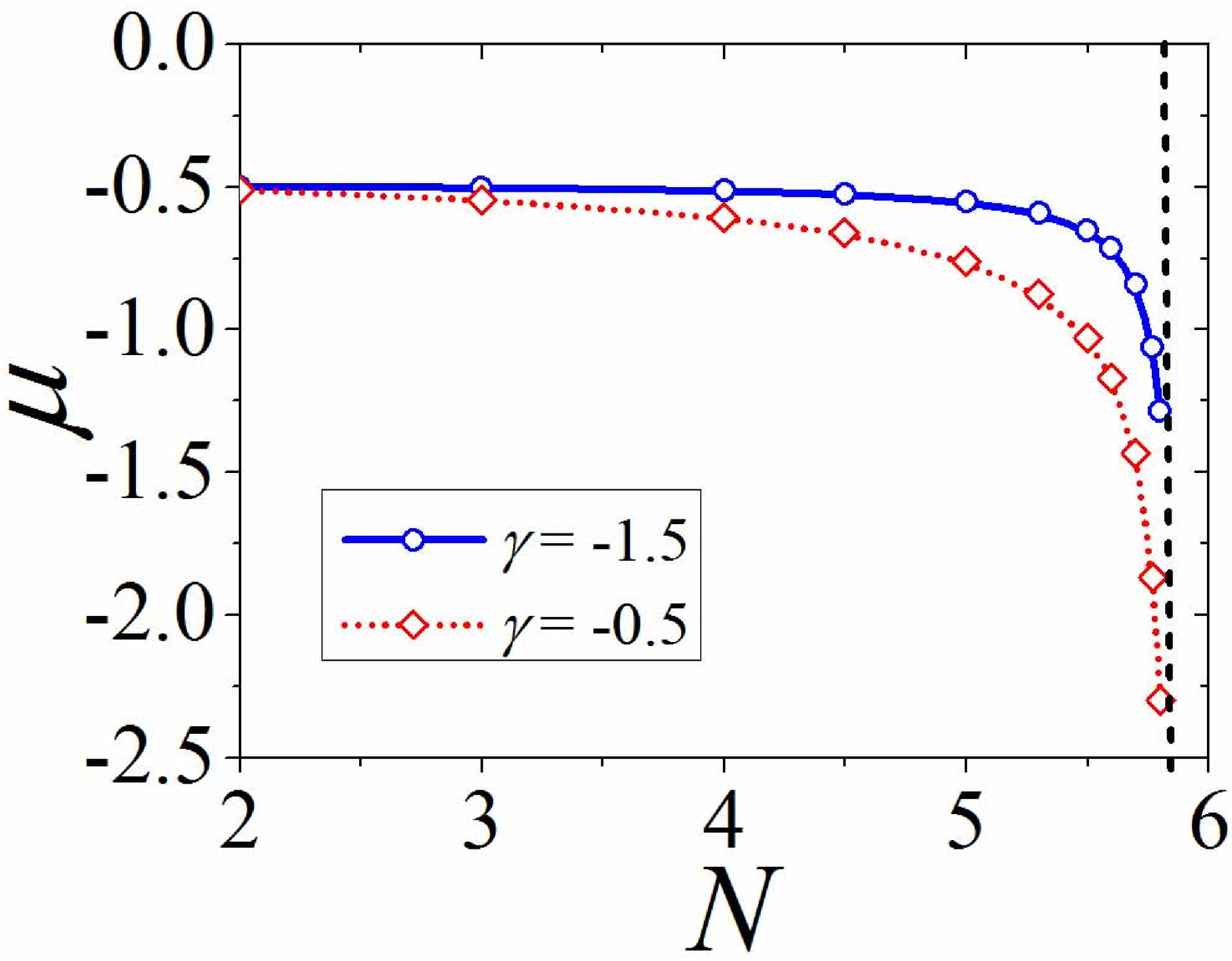}} %
\subfigure[]{\includegraphics[width=0.3\columnwidth]{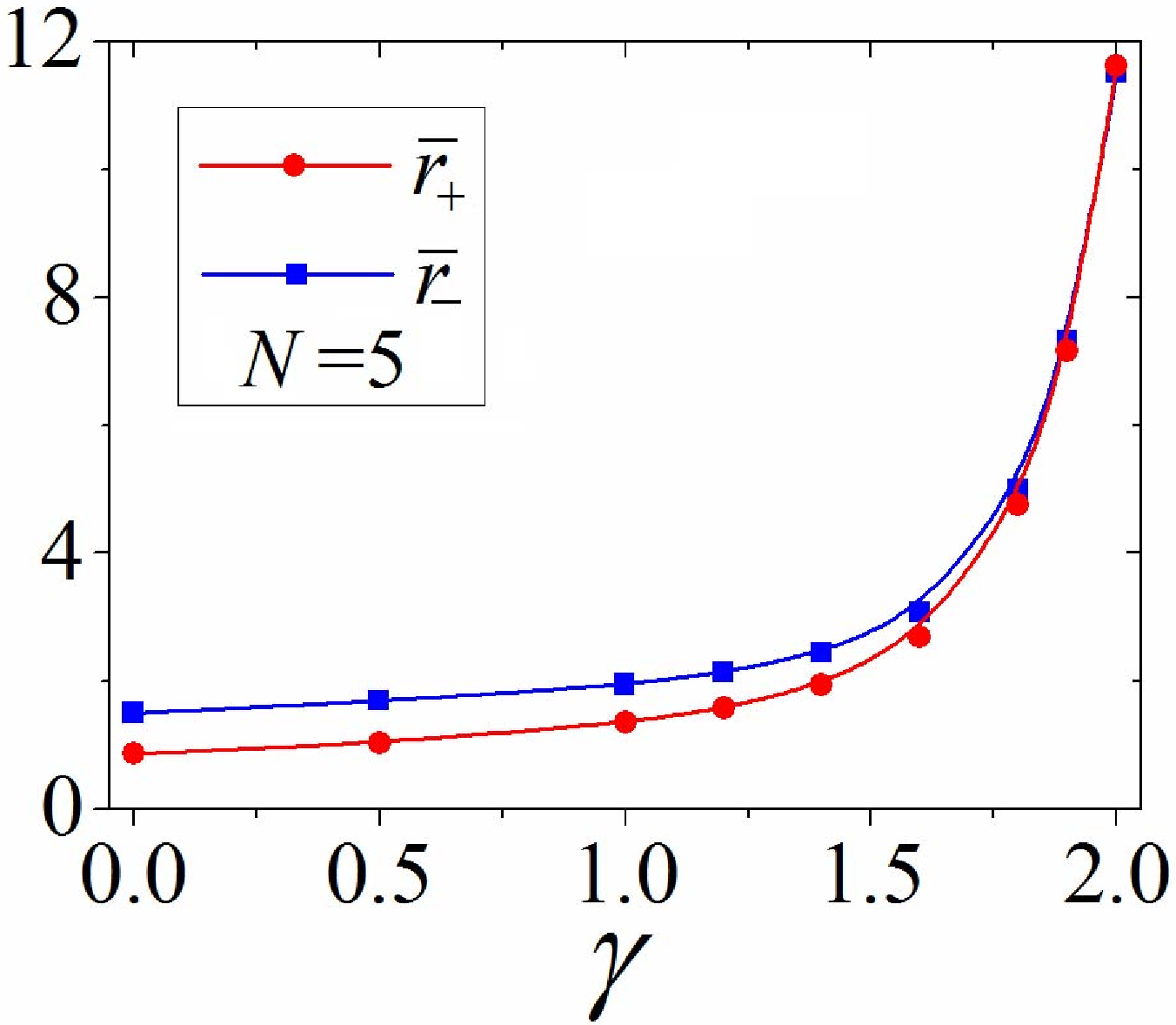}} %
\subfigure[]{\includegraphics[width=0.3\columnwidth]{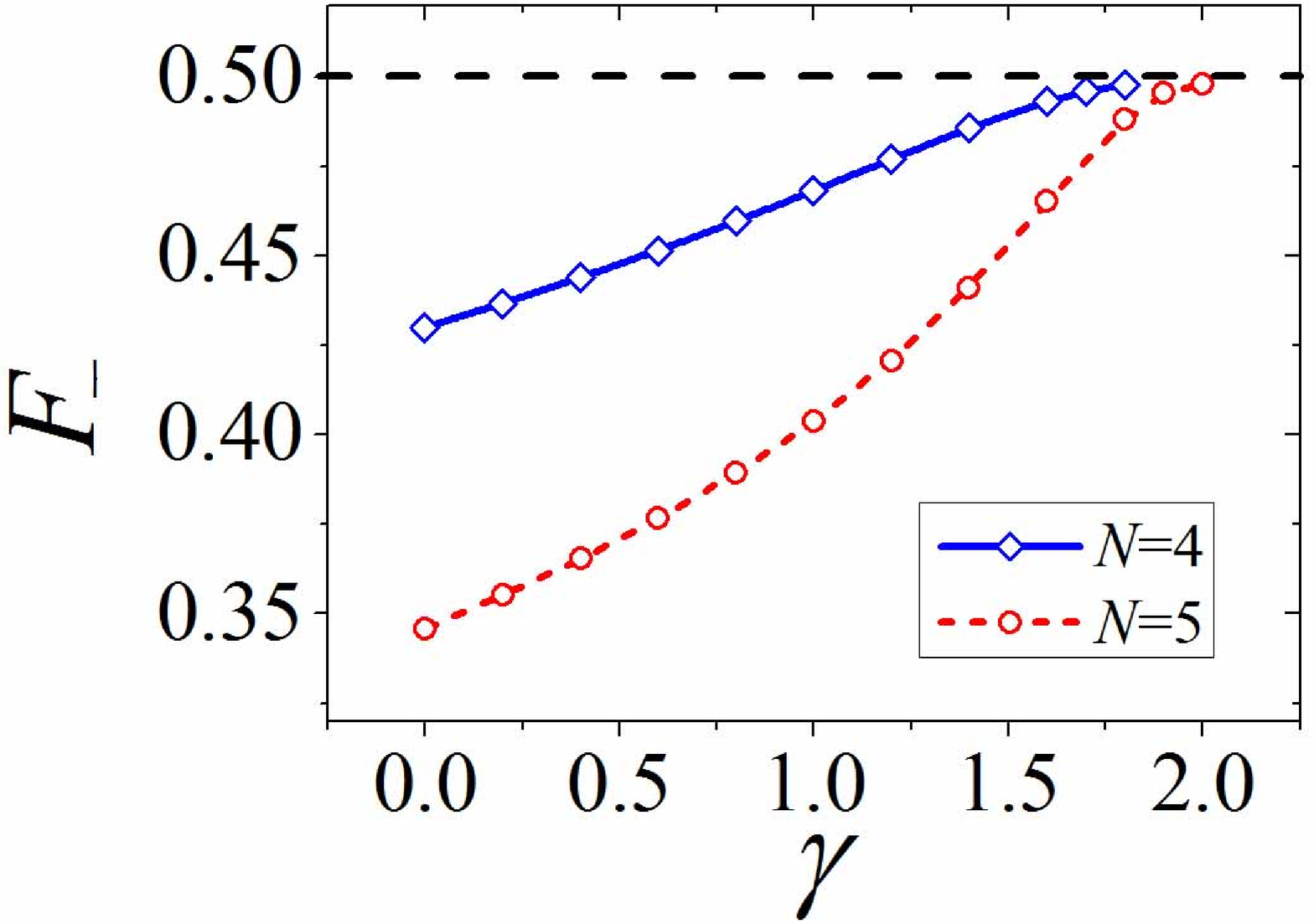}} %
\subfigure[]{\includegraphics[width=0.3\columnwidth]{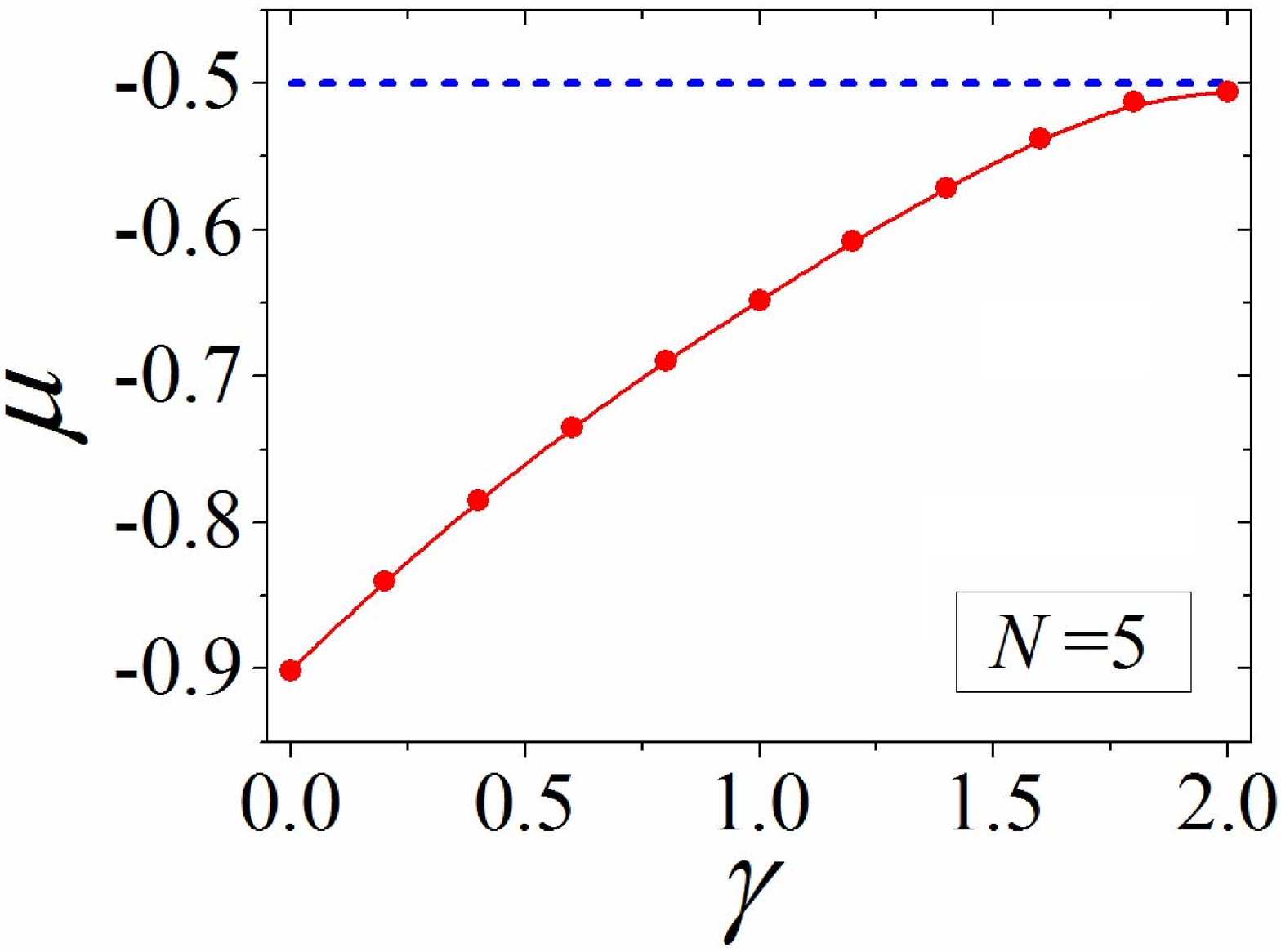}} %
\subfigure[]{\includegraphics[width=0.3\columnwidth]{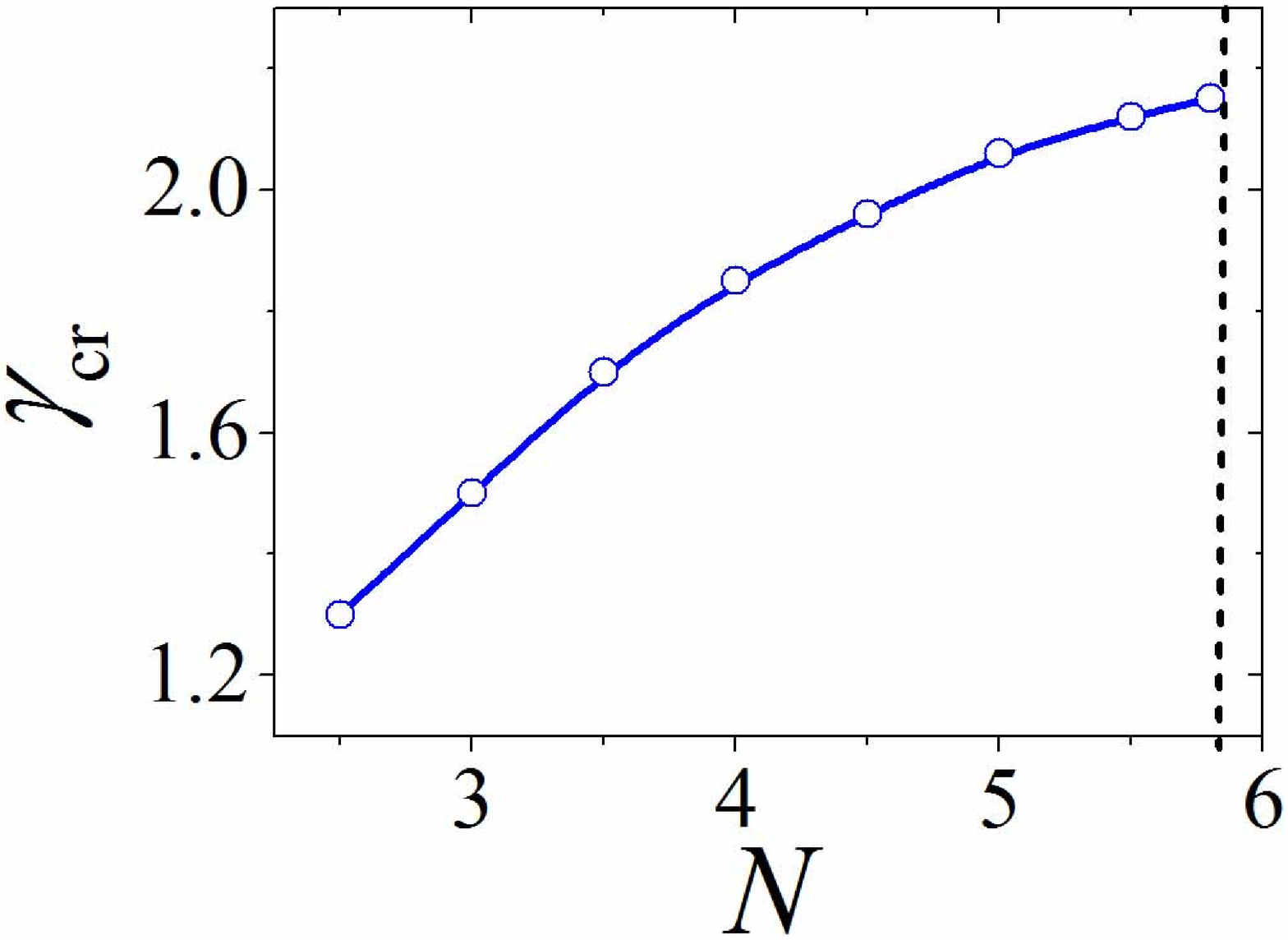}}
\caption{(a) Typical examples of stable SVs for $\protect\gamma =-1$ and $%
-1.5$ (top and bottom panels, respectively). (b) The chemical potential of
the SV versus $N$ for $\protect\gamma =-1.5$ and $-0.5$ (the blue solid and
red dashed curves, respectively). The vertical dashed line designates the
norm of the Townes soliton, $N_{\mathrm{Townes}}=5.85$. (c) Average radii of
the two components (with $N=5$), $\bar{r}_{\pm }$, versus $\protect\gamma $
[see Eq. (\protect\ref{radii})]. (d) The norm share of the vortex component
versus $\protect\gamma $ for $N=4$ and $5$ (the blue solid and red dashed
curves, respectively). The horizontal dashed line corresponds to $F_{-}=0.5$%
. (e) The chemical potential of the SV with $N=5$ versus $\protect\gamma $.
The horizontal dashed line designates the limit of the infinitely broad mode
with $\protect\gamma =-1/2$. (f) $\protect\gamma _{\mathrm{cr}}$ versus $N$.
In all the panels the SPM coefficient is fixed as $g=-1$.}
\label{SemiV}
\end{figure}

In the system with the same signs of the competing SPM (repulsive) and XPM
(attractive) terms as above, i.e., $g,\gamma >0$, imaginary-time simulations
of Eq. (\ref{GPE}), starting from the SV ansatz (\ref{SVS}), converge to the
MM solution, clearly demonstrating that the system supports no SVs in this
case. SV solutions are readily generated in the opposite case, with
attractive SPM and repulsive XPM terms, i.e., $g,\gamma <0$ in Eq. (\ref{GPE}%
). Fixing in this case, as said above, $g=-1$ by means of rescaling, we find
stable SVs with the total norm taking values in the natural interval of $%
N<N_{\mathrm{Townes}}\approx 5.85$. Typical examples of numerical solutions
for stable SVs are displayed in Fig. \ref{SemiV}(a), which shows that the
size of the SV increases with the increase of $|\gamma |$. The stability of
the SVs is corroborated by the fact that their $\mu (N)$ dependence also
satisfies the above-mentioned VK criterion, $d\mu /dN<0$, as seen in Fig. %
\ref{SemiV}(b).

To highlight a relation between the size and structure of the SV and the
magnitude of the repulsive XPM coefficient $|\gamma |$, which competes with
the attractive SPM, we define effective widths of the two components and the
share of the norm in the vortex component, respectively, as
\begin{equation}
\bar{r}_{\pm }=\left( {\frac{\int_{0}^{\infty }|\phi _{\pm }(r)|^{2}r^{3}dr}{%
\int_{0}^{\infty }|\phi _{\pm }(r)|^{2}rdr}}\right) ^{1/2},\quad {F_{-}={%
\frac{N_{-}}{N}}}.  \label{radii}
\end{equation}%
Figure \ref{SemiV}(c) displays the widths as functions of $\gamma $ for $%
(N,g)=(5,-1)$ (naturally, the width is larger for the vortical component, $%
\phi _{-}$, which has the ``hole" at its center). The norm share of the
vortex component is displayed, as a function of $\gamma $, in Fig. \ref%
{SemiV}(d) for $(N,g)=(4,-1)$ and $(5,-1)$. For the SVs, unlike the MMs,
point $|g|=|\gamma |=1$ is not a critical one, as norms of the two SV's
components are strongly different. However, Figs. \ref{SemiV}(c,d)
demonstrates that there is a critical value $\gamma _{\mathrm{cr}}$ of $%
\gamma $, with the widths diverging and the norm share of the vortex
component, $F_{-}$, approaching $0.5$ at $\gamma \rightarrow \gamma _{%
\mathrm{cr}}$. In the same limit, the chemical potential, $\mu $,\
approaches the above-mentioned limit value, $\mu =-1/2$ [see a typical
example in Fig. \ref{SemiV}(e) for $N=5$], i.e., at $\gamma =\gamma _{%
\mathrm{cr}}$ the SV family displayed in Fig. \ref{SemiV} degenerates into
an infinitely extended state with an infinitesimal amplitude. Obviously, $%
\gamma _{\mathrm{cr}}$ is a function of $N$, with $\gamma _{\mathrm{cr}%
}\rightarrow 0$ in the linear limit, i.e., at $N\rightarrow 0$. Figure \ref%
{SemiV}(f) displays $\gamma _{\mathrm{cr}}$ in the interval of $3<N<N_{%
\mathrm{Townes}}$, the maximum value being $\gamma _{\mathrm{cr}%
}(N\rightarrow N_{\mathrm{Townes}})\approx 2.15$.

\section{Mixed modes under the action of the Lee-Huang-Yang terms.}

\begin{figure}[t]
{\includegraphics[width=0.7\columnwidth]{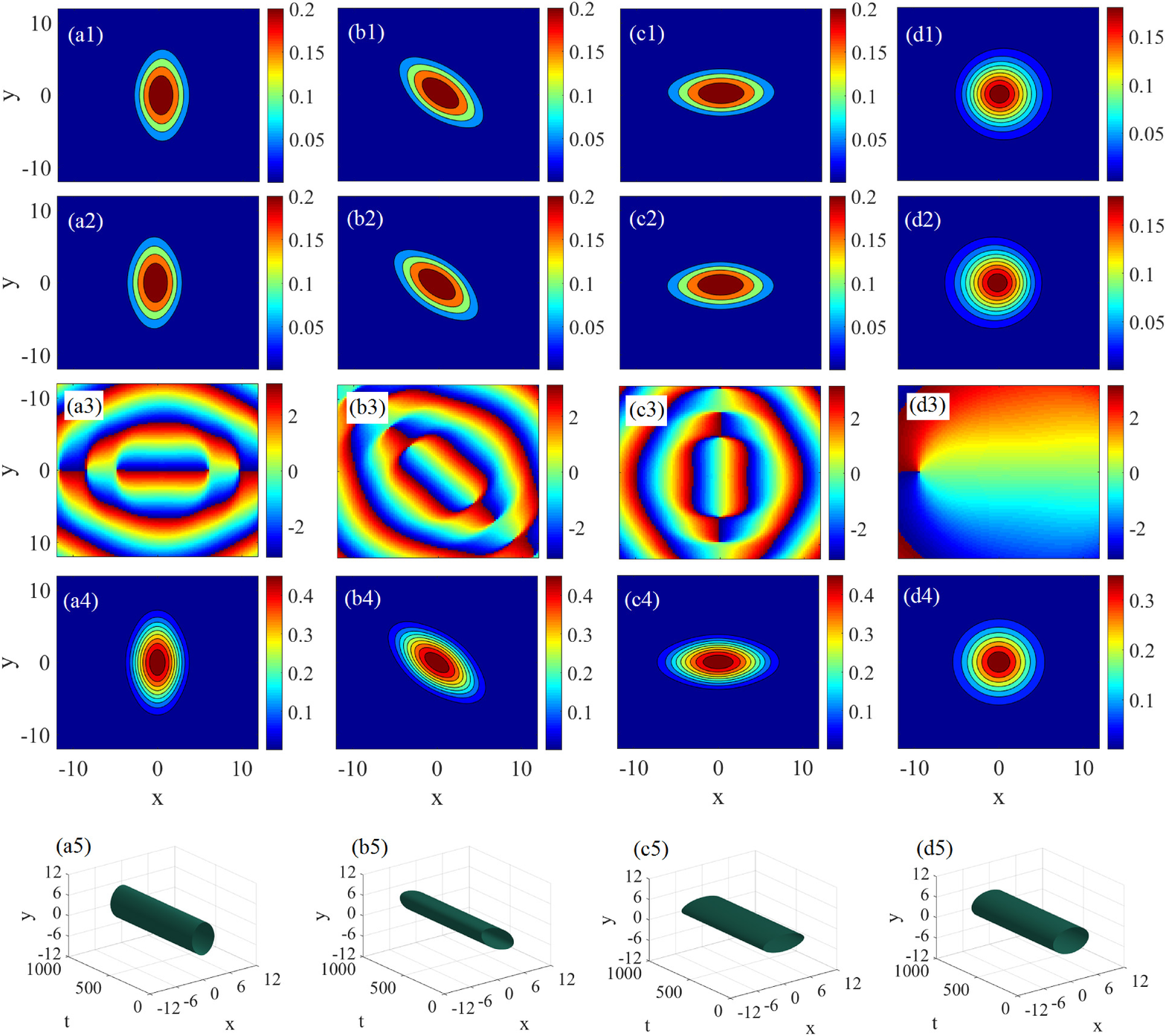}}
\caption{Column (a-c) (from left to right) are typical examples of quantum
droplets with $(g,N,\protect\lambda )=(2,20,1)$ with the vertical (a1-a5),
diagonal (b1-b5) and horizontal (c1-c5) orientations, respectively. The
first and second rows display density patterns of the two components, $|%
\protect\phi _{+}(\mathbf{r})|^{2}$ and  $|\protect\phi _{-}(\mathbf{r})|^{2}
$, respectively,panels in  the third row are phase patterns of $\protect\phi %
_{+}$, fourth row shows the total density profile, $n(\mathbf{r})=|\protect%
\phi _{+}(\mathbf{r})|^{2}+|\protect\phi _{-}(\mathbf{r})|^{2}$, and the
final row illustrates the full stability of the quantum droplets in direct
simulations of Eq. (\protect\ref{GPE2}) [shown by the evolution of $n(%
\mathbf{r},t)$]. The last column (d) is a typical example of the quantum
droplet with $(g,N,\protect\lambda )=(2,20,0.2)$, which is nearly isotropic
when the SOC$\ $strength, $\protect\lambda $, is small enough. The meaning
of different rows in this column are same as in the others.}
\label{QDexp}
\end{figure}
\begin{figure}[t]
{\includegraphics[width=0.7\columnwidth]{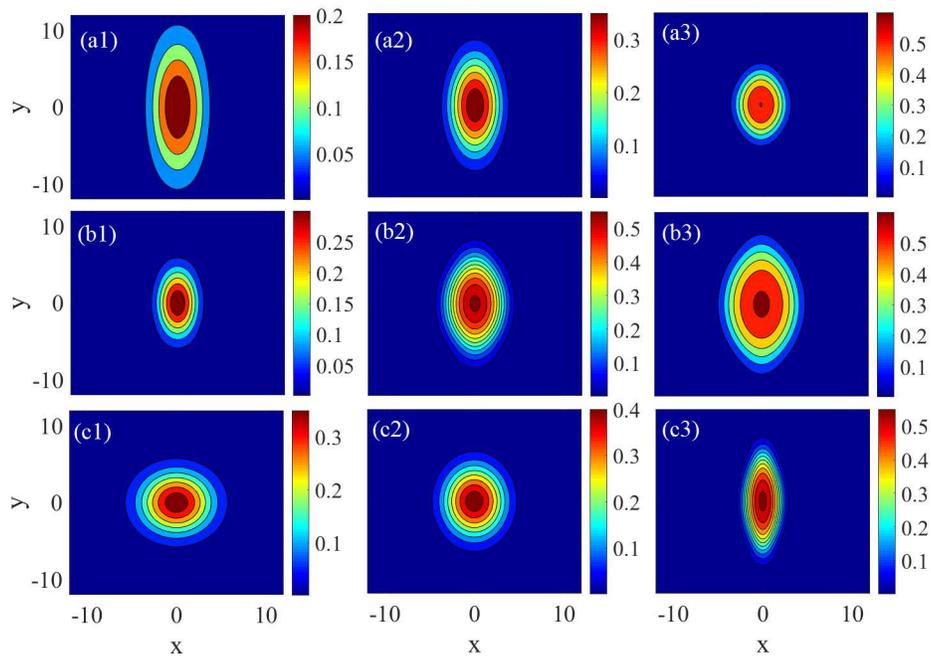}}
\caption{The variation of the total-density profile, $n(\mathbf{r})=|\protect%
\phi _{+}(\mathbf{r})|^{2}+|\protect\phi _{-}(\mathbf{r})|^{2}$, at diferent
values of $(g,N,\protect\lambda )$. The first row: $g=1$ in (a1), $2$ in
(a2), and $3$ in (a3). In this row, $N=20$ and $\protect\lambda =1$ are
fixed. The second row: $N=10$ in (b1), $30$ in (b2), and $50$ in (b3).In
this row, $g=2$ and $\protect\lambda =1$ are fixed. The third row:  $\protect%
\lambda =0$ in (c1), $0.5$ in (c2), and $2$ in (c3). In this row, $g=2$ and $%
N=20$ are fixed. }
\label{QDexp2}
\end{figure}

\begin{figure}[t]
{\includegraphics[width=0.7\columnwidth]{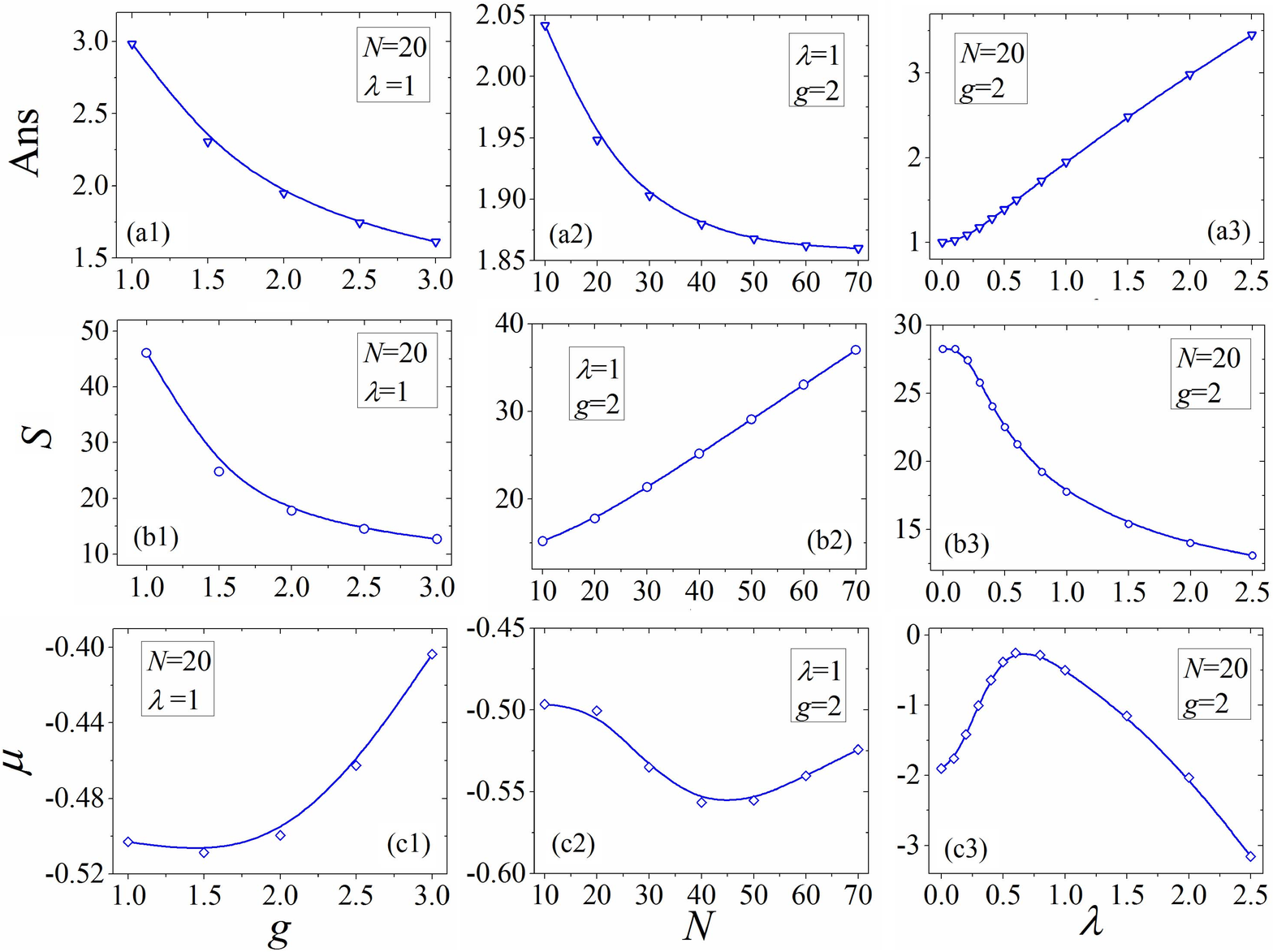}}
\caption{The first row: the anisotropy factor of the quantum droplet, $%
\mathrm{Ans}$, versus $g$ (a1), $N$ (a2) and $\protect\lambda $ (a3), see
Eq. (\protect\ref{Ans-S}) for its definition. The second row: the effective
area, $S$, versus $g$ (a1), $N$ (a2) and $\protect\lambda $ (a3). The third
row: the chemical potential, $\protect\mu $, versus $g$ (c1), $N$ (c2) and $%
\protect\lambda $ (c3). }
\label{charQD}
\end{figure}

In the case of the repulsive SPM, the LHY theory, which introduces effects
of quantum fluctuations around the mean-field state, gives rise to
post-mean-field corrections which are crucially important for the
stabilization of self-trapped QDs by means of the effective higher-order
self-repulsion \cite{Petrov2015}-\cite{Baillie2016}. In particular, it
removes the collapse induced by the attractive XPM at $N=N_{\mathrm{thr}}$,
see Eq. (\ref{th}). It is relevant to mention that the LHY correction to the
mean-field energy is itself affected by the presence of SOC in the spinor
BEC, due to the degeneracy of atomic zero-point-energy spectrum imposed by
the SOC, as shown in Ref. \cite{Zhengwei2013}. As a result, the interplay
with SOC gives rise to an effective renormalization of the strength of the
original LHY correction, in the case of moderately strong SOC. If the SOC is
too strong, Fig. 6 in Ref. \cite{Zhengwei2013} suggests that the LHY
correction may qualitatively change its form. In fact, the mean-field part
of the model may also undergo a dramatic change in the case of strong SOC,
replacing the regular solitons by gap solitons, as shown in Ref. \cite%
{we2017}. Thus, both the mean-field and LHY\ parts of the model adopted in
the present work are relevant (the LHY part is displayed below), provided
that the SOC does not grow too strong.

Further, the model assumes effective locality of the LHY correction to the
mean-field energy. This assumption is definitely corroborated, for the
nondipolar binary condensates, by the previous theoretical results \cite%
{Petrov2015,QDterm,Sadhan-nondip}, as well as by the very recent
experimental work \cite{Leticia} reporting the creation of stable QDs
supported by purely contact (local) interaction in the gas of $^{39}$K
atoms.

According to Ref. \cite{QDterm}, the LHY terms, which are added to the
coupled GPEs, are derived from the expression for the LHY energy. In the
general case, this expression is quite complex, as shown by Eqs. (4) and (3)
in Ref. \cite{QDterm}. However, it is shown in the same work that, in the
case of
\begin{equation}
n_{\mathrm{2D}}\ll a^{-2},  \label{n2D}
\end{equation}%
where $n_{\mathrm{2D}}$ is the peak value of the two-dimensional atomic
density and $a$ is the scattering length responsible for the self-repulsion
of each component, the energy can be cast in an essentially simpler form
given by Eqs. (5) and (6) in Ref. \cite{QDterm}. A straightforward estimate
demonstrates that inequality (\ref{n2D}) amounts to the following condition
for a characteristic size $l_{\mathrm{2D}}$ of a localized 2D pattern:%
\begin{equation}
l_{\mathrm{2D}}^{2}\gg aa_{\perp },  \label{2D}
\end{equation}%
where $a_{\perp }$ is the length of the transverse confinement of the
quasi-2D condensate. Condition (\ref{2D}) definitely holds for any
non-collapsing soliton. In this case, the nonlinear terms in the coupled
GPEs can be derived from the respective part of the system's energy density,
which was obtained in Ref. \cite{QDterm} (it effectively represents
zero-point energy of the Bogoliubov modes on top of the mean-field state):%
\begin{equation}
\mathcal{E}_{\mathrm{2D}}=\frac{1}{2}g\left( \left\vert \phi _{+}\right\vert
^{2}-\left\vert \phi _{-}\right\vert ^{2}\right) ^{2}+\frac{g^{2}}{8\pi }%
\left( \left\vert \phi _{+}\right\vert ^{2}+\left\vert \phi _{-}\right\vert
^{2}\right) ^{2}\ln \left( \frac{\left\vert \phi _{+}\right\vert
^{2}+\left\vert \phi _{-}\right\vert ^{2}}{\sqrt{e}n_{0}}\right) ,
\label{density}
\end{equation}%
where $n_{0}\equiv |\phi^{f} _{+}| ^{2}+|\phi^{f}_{-}| ^{2}$ is the density of the symmetric ($|\phi^{f}_{+}|^{2}=|\phi^{f} _{-}|^{2}$) flat state
corresponding to this expression, and $e$ is the base of the natural
logarithms.

Finally, coupled GPEs in the LHY form are derived by the variation of the
total energy corresponding to density (\ref{density}), as 
\begin{eqnarray}
i\frac{\partial \phi _{+}}{\partial t} &=&-\frac{1}{2}\nabla ^{2}\phi
_{+}+\lambda \hat{D}\phi _{-}+g(|\phi _{+}|^{2}-|\phi _{-}|^{2})\phi _{+}+%
\frac{g^{2}}{4\pi }\left( |\phi _{+}|^{2}+|\phi _{-}|^{2}\right) \ln \left(
|\phi _{+}|^{2}+|\phi _{-}|^{2}\right) \phi _{+},  \notag \\
i\frac{\partial \phi _{-}}{\partial t} &=&-\frac{1}{2}\nabla ^{2}\phi
_{-}-\lambda \hat{D}^{\ast }\phi _{+}+g(|\phi _{-}|^{2}-|\phi _{+}|^{2})\phi
_{-}+\frac{g^{2}}{4\pi }\left( |\phi _{+}|^{2}+|\phi _{-}|^{2}\right) \ln
\left( |\phi _{+}|^{2}+|\phi _{-}|^{2}\right) \phi _{-},  \label{GPE2}
\end{eqnarray}%
where the remaining scaling invariance is used to fix $n_{0}\equiv 1$.

Numerical solutions of Eq. (\ref{GPE2}) of the MM type, which are relevant
here because of the opposite signs of the self- and cross-interactions in
the terms $\sim g$, were produced by means of the same techniques as used
above for the solution of Eq. (\ref{GPE}). Thus, the controlled parameters
in this section are a set of $(g,N,\lambda )$.

Numerical solution of Eqs. (\ref{GPE2}) produces anisotropic (elongated) QD
density profiles when $\lambda \neq 0$. The size and area of the QDs are
much larger than for the solitons found in above sections. In accordance
with Eq. (\ref{rot}), in the presence of the Rashba-type SOC, the
anisotropic soliton may be oriented in any direction in the $\left(
x,y\right) $ plane. To illustrate this property, typical examples of the QDs
with vertical, diagonal, and horizontal directions are displayed in the
first three columns of Fig. \ref{QDexp}. When $\lambda $ is smaller enough,
the QD\ profile is nearly isotropic, see an example in the last column of
Fig. \ref{QDexp}. In the fourth row of Fig. \ref{QDexp}, one can see that
the total-density profile, $n(\mathbf{r})=|\phi _{+}(\mathbf{r})|^{2}+|\phi
_{-}(\mathbf{r})|^{2}$, naturally has the oval shape similar to that of each
component.

For the comparison of characteristics of the QD with different sets of
control parameters, we plot the patterns of $n(\mathbf{r})$ corresponding to
different $(g,N,\lambda )$ sets in Fig. \ref{QDexp2}, choosing the vertical
orientation for all the QDs. The figure shows that the area and the
anisotropy of the $n(\mathbf{r})$ patterns strongly depend on the parameters.

In order to further quantify the dependence of the QDs on controlled
parameters, we define average horizontal and vertical widths for $n(\mathbf{r%
})$ as
\begin{equation*}
\{W_{x},W_{y}\}=\left( {\frac{1}{N}}\int \int \{x^{2},y^{2}\}n(\mathbf{r}%
)dxdy\right) ^{1/2}.
\end{equation*}%
The degree of the anisotropy and the area of the QD can be defined by $%
W_{x,y}$ as
\begin{equation}
\mathrm{Ans}=W_{y}/W_{x},\quad S=\pi W_{x}\times W_{y}.  \label{Ans-S}
\end{equation}%
These two characteristics, as well as the chemical potential of the QD, are
displayed, as functions of $g$, $N$ and $\lambda $ in Fig. \ref{charQD}. It
clearly demonstrates that: (i) the increase of the nonlinearity strength, $g$%
, reduces the anisotropy and effective area of the QD; (ii) the increase of
the total norm, $N$, reduces the anisotropy and increases the effective,
respectively; (iii) opposite to the effect of $N$, the increase of the SOC
strength, $\lambda $, can increase the anisotropy and decreases the
effective area, respectively.

It is also relevant to stress that, unlike the first and second row of Fig. %
\ref{charQD}, in the last row of Fig. \ref{charQD} the dependence of the
chemical potential, $\mu $, on parameters $(g,N,\lambda )$ is not
monotonous. In particular, the non-monotonous dependence $\mu (N)$ implies
that the QD family does not satisfy the VK criterion, $d\mu /dN<0$, although
the entire QD family, including the segment with $d\mu /dN>0$, is found to
be \emph{completely stable}. The violation of the VK criterion as the
necessary stability condition \cite{VaKo}-\cite{Fibich} is explained by the
fact that the QD modes are supported not by the purely attractive
interaction between the two components, but actually by its combination with
the self-repulsive LHY terms.

\section{Quantum droplets under the action of mixed Rashba-Dresselhaus SOC}

\begin{figure}[t]
{\includegraphics[width=0.8\columnwidth]{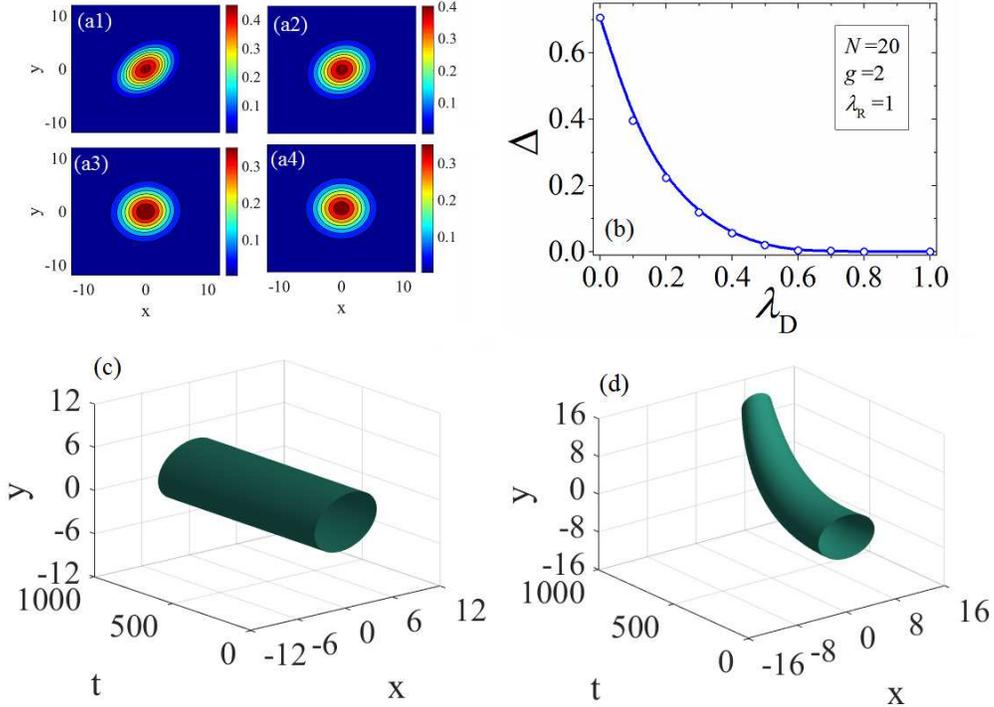}}
\caption{(a1-a4) Density patterns of the total density, $n(\mathbf{r})$ with
the strength of the Dresselhaus SOC interaction $\protect\lambda _{\mathrm{D}%
}=0.1$ (a1), $0.3$ (a2), $0.5$ (a3) and $1$ (a4). (b) The diagonal
anisotropy, $\Delta $ [see Eq. (\protect\ref{Delta}) for its definition] as a
function of $\protect\lambda _{D}$. (c,d) Direct simulations of the
evolution of the quantum droplets, shown by means of $n(\mathbf{r},t)$, with
$\protect\lambda _{\mathrm{D}}=0.2$ and $1$, respectively (the anisotropic
droplet is stable, while the isotropic one is not). In all panels, values $%
(g,N,\protect\lambda _{\mathrm{R}})=\left( 2,20,1\right) $ are fixed.}
\label{RD_QD}
\end{figure}

As shown in Ref. \cite{SVS3}, the addition of the Dresselhaus terms to the
SOC\ dominated by the Rashba coupling tends to delocalize the solitons in
the framework of the mean-filed GPE system (which does not include the LHY
terms). Therefore, the existence of stable 2D solitons under the action of
the mixed Rashba-Dresselhaus SOC is a challenging issue. In this case, Eq. (%
\ref{GPE2}) is rewritten as
\begin{eqnarray}
i\frac{\partial \phi _{+}}{\partial t} &=&-\frac{1}{2}\nabla ^{2}\phi
_{+}+\left( \lambda _{\mathrm{R}}\hat{D}-i\lambda _{\mathrm{D}}\hat{D}^{\ast
}\right) \phi _{-}+g(|\phi _{+}|^{2}-|\phi _{-}|^{2})\phi _{+}+\frac{g^{2}}{%
4\pi }\left( |\phi _{+}|^{2}+|\phi _{-}|^{2}\right) \ln \left( |\phi
_{+}|^{2}+|\phi _{-}|^{2}\right) \phi _{+},  \notag \\
i\frac{\partial \phi _{-}}{\partial t} &=&-\frac{1}{2}\nabla ^{2}\phi
_{-}-\left( \lambda _{\mathrm{R}}\hat{D}^{\ast }+i\lambda _{\mathrm{D}}\hat{D%
}\right) \phi _{+}+g(|\phi _{-}|^{2}-|\phi _{+}|^{2})\phi _{-}+\frac{g^{2}}{%
4\pi }\left( |\phi _{+}|^{2}+|\phi _{-}|^{2}\right) \ln \left( |\phi
_{+}|^{2}+|\phi _{-}|^{2}\right) \phi _{-},  \label{GPE3}
\end{eqnarray}%
where $\lambda _{\mathrm{R}}$ and $\lambda _{\mathrm{D}}$ are strengths of
the Rashba and Dresselhaus SOC. Note that the Dresselhaus-only system, with $%
\lambda _{\mathrm{R}}=0$ and $\lambda _{\mathrm{D}}\neq 0$, is made
tantamount to its Rashba-only counterpart, with $\lambda _{\mathrm{R}}\neq 0$
and $\lambda _{\mathrm{D}}=0$, by swapping $x\leftrightarrow y$ \cite%
{Sun2017}. On the other hand, the mixed form of the SOC operators in Eq. (%
\ref{GPE3}) obviously breaks the rotational symmetry defined above by Eq. (%
\ref{rot}).

It was found in Ref. \cite{SVS3} that, for fixed $\lambda _{R}$, there are
critical values, $\left( \lambda _{D}^{\mathrm{cr}}\right) _{\mathrm{SV}%
}<\left( \lambda _{D}^{\mathrm{cr}}\right) _{\mathrm{MM}}<\lambda _{R}$, of
the Dresselhaus-SOC strength, above which the SVs and MMs, respectively,
disappear through delocalization. In particular, this implies that the
mean-field solitons never exist at $\lambda _{D}=\lambda _{R}$, i.e., for
equal strengths of both SOC interactions.

In this section we aim to briefly explore the possibility of the existence
of stable QDs in the framework of Eq. (\ref{GPE3}), which includes the LHY
terms, in the case of different strengths of the Rashba and Dresselhaus
interactions. Numerical solutions demonstrate that, in this LHY-affected
system, QDs can be found with \emph{arbitrary} values of $\lambda _{\mathrm{R%
}}$ and $\lambda _{\mathrm{R}}$. To illustrate these findings, we currently
fix $\lambda _{\mathrm{R}}=1$ and vary $\lambda _{\mathrm{D}}$ in the
interval of $[0,1]$. Typical examples of the the respective numerical
solutions for QD with different values of $\lambda _{\mathrm{D}}$ and fixed
values of $(g,N)$ are shown in Fig. \ref{RD_QD}(a1-a4). In particular, the
orientation of the QDs tends to be diagonal. In the limit of $\lambda _{%
\mathrm{D}}=\lambda _{\mathrm{R}}$, the QDs do not suffer delocalization,
unlike their mean-field counterparts (see above), but rather become
isotropic. To quantify the latter property, we define the diagonal
anisotropy, in terms of the total density of the two components, $n(\mathbf{r%
})$, as
\begin{equation}
\Delta ={\frac{1}{N}}\int \int |n(x,y)-n(-y,x)|dxdy,  \label{Delta}
\end{equation}%
where $n(-y,x)$ is the $90^{\circ}$ counterclockwise rotation of $n(x,y)$. Figure \ref{RD_QD}(b) shows $\Delta $ as a function of $\lambda _{\mathrm{D}}
$ for fixed values of other parameters, $(g,N,\lambda _{\mathrm{R}})=(2,20,1)
$. In Fig. \ref{RD_QD}, the QD profile become practically isotropic at $%
\lambda _{\mathrm{D}}>0.6$.

Lastly, direct simulations of the QD evolution in the framework of Eq. (\ref%
{GPE3}) demonstrate that asymmetric QDs are stable [see Fig. \ref{RD_QD}%
(c)], while effectively isotropic ones tend to be unstable against
spontaneous drift in the diagonal direction, as shown in Fig. \ref{RD_QD}(d).

\section{Conclusion}

The first objective of this work is to study 2D matter-wave solitons in SOC
(spin-orbit-coupled) two-component BEC with opposite signs of the cubic SPM
and XPM interactions, unlike the recently studied system in which stable MM
(mixed-mode) and SV (semi-vortex) solitons are created by entirely
attractive interactions. In the present case, stable MMs were found provided
that the XPM attraction is stronger than the SPM repulsion. These modes
exist with the total norm falling below a critical value corresponding to
the onset of the collapse. Excited states of the MMs were found too, being
completely unstable (in some cases, their instability may be weak). In the
opposite case of the attractive SPM and repulsive XPM, the system supports
stable SVs with the total norm $N<N_{\mathrm{Townes}}$, where $N_{\mathrm{%
Townes}}$ is the well-known critical norm corresponding to the Townes
solitons.

Another setting studied in this work is one with the repulsive SPM and
attractive XPM interactions, including the post-mean-field LHY
(Lee-Huang-Yang) terms, which eliminate the collapse and help to create
stable QDs (quantum drops) of the MM type with arbitrarily large values of
the norm. In particular, the QDs feature a large size and an anisotropic
(elliptical) density profile, that may be oriented in any direction, when
the LHY terms play an essential role, and the SOC is taken in the Rashba
form. The entire family of the QDs is stable in this case, even if it does
not satisfy the Vakhitov-Kolokolov criterion. The growth of the SOC strength
leads to the increase the anisotropy degree of the QD and decrease of its
size. The QD family remains partly stable under the action of the mixed
Rashba-Dresselhaus SOC.

A challenging possibility is to extend the present analysis to 3D systems.
It was recently found that SOC can support metastable solitons of both SV
and MM types in the 3D spinor BEC with fully attractive cubic interactions
\cite{Yongchang}. It may be relevant to consider 3D system with the
competing signs of the SPM and XPM interactions, as well as MMs in 3D
stabilized by the respective LHY terms.

\begin{acknowledgments}
We appreciate valuable discussions with G. E. Astrakharchik, H. Sakaguchi,
E. Ya. Sherman, Zheng Wei, and Pang Wei. This work was supported by NNSFC
(China) through Grant No. 11575063, 61471123, 61575041, by the Natural
Science Foundation of Guangdong Province, through Grant No. 2015A030313639,
by the joint program in physics between NSF and Binational (US-Israel)
Science Foundation through project No. 2015616, and by the Israel Science
Foundation through Grant No. 1286/17. B.A.M. appreciates a foreign-expert
grant of the Guangdong province (China).
\end{acknowledgments}


\begin{thebibliography}{99}

\bibitem{Segev1994} Segev M, Valley G C, Crosignani B, DiPorto P, and Yariv
A, 1994 \emph{Phys. Rev. Lett.} \textbf{73}, 3211.

\bibitem{Mihalache2006} Mihalache D, Mazilu D, Malomed B A, Lederer F,
Crasovan L -C, Kartashov Y V, and Torner L 2006 \emph{Phys. Rev. E} \textbf{%
74}, 047601.

\bibitem{Mihalache22006} Mihalache D, Mazilu D, Lederer F, Kartashov Y V,
Crasovan L -C, Torner L, and Malomed B A 2006 \emph{Phys. Rev. Lett.}
\textbf{97}, 073904.

\bibitem{Mihalache32006} Mihalache D, Mazilu D, Lederer F, Malomed B A,
Kartashov Y V, Crasovan L -C, and Torner L 2006 \emph{Phys. Rev. E} \textbf{%
73}, 025601(R).

\bibitem{Skupin} Skupin S, Bang O, Edmundson D, and Kr\'{o}likowski W 2006
\emph{Phys. Rev. E} \textbf{73}, 066603.

\bibitem{Pedri2005} Pedri P, and Santos L 2005 \emph{Phys. Rev. Lett.}
\textbf{95} 200404.

\bibitem{Nath2008} Nath R, Pedri P, and Santos L 2008 \emph{Phys. Rev. Lett.}
\textbf{101}, 210402.

\bibitem{Tikhonenkov2008} Tikhonenkov I, Malomed B A, and Vardi A 2008 \emph{%
Phys. Rev. A} \textbf{78}, 043614.

\bibitem{Yongyao2013} Li Y, Liu J, Pang W, and Malomed B A 2013 \emph{Phys.
Rev. A} \textbf{88}, 053630.

\bibitem{Tikhonenkov22008} Tikhonenkov I, Malomed B A, and Vardi A 2008
\emph{Phys. Rev. Lett.} \textbf{100}, 090406.

\bibitem{Raghunandan2015} Raghunandan M, Mishra C, Lakomy K, Pedri P, Santos
L, and Nath R 2015 \emph{Phys. Rev. A} \textbf{92}, 013637.

\bibitem{Jiasheng} Huang J, Jiang X, Chen H, Fan Z, Pang W, Li Y 2015 \emph{%
Front. Phys.} \textbf{10}, 100507. %

\bibitem{SVS1} Sakaguchi H, Li B, Malomed B A 2014 \emph{Phys. Rev. E}
\textbf{89}, 032920.

\bibitem{SVS2} Jiang X, Fan Z, Chen Z, Pang W, Li Y, and Malomed B A 2016
\emph{Phys. Rev. A}, \textbf{93}, 023633.

\bibitem{Bingjin2017} Liao B, Li S, Huang C, Luo Z, Pang W, Tan H, Malomed B
A, Li Y 2017 \emph{Phys. Rev. A} \textbf{96}, 043613.

\bibitem{SVS3} Sakaguchi H, Sherman E Y, and Malomed B A 2016 \emph{Phys.
Rev. E} \textbf{90}, 032202.

\bibitem{SVSNJP} Sakaguchi H, Malomed B A 2016 \emph{New J. Phys.} \textbf{18%
} 105005.

\bibitem{SVS4} Gautam S, and Adhikari S K 2017 \emph{Phys. Rev. A} \textbf{95%
}, 013608.

\bibitem{Guihua2017} Chen G, Liu Y, Wang H 2017 \emph{Commun. Nonlinear Sci.
Numer. Simulat.} \textbf{48}, 318.

\bibitem{Yongchang} Zhang Y, Zhou Z, Malomed B A, and Pu H 2015 \emph{Phys.
Rev. Lett.} \textbf{115}, 253902.


\bibitem{Yongping} Zhang Y, Mossman M E, Busch T, Engels P, and Zhang C 2016
\emph{Front. Phys.} \textbf{11}, 118103.


\bibitem{we2017} Li Y, Liu Y, Fan Z, Pang W, Fu S, and Malomed B A 2017
\emph{Phys. Rev. A} \textbf{95}, 063613.


\bibitem{SOCoptics} Kartashov Y V, Malomed B A, Konotop V V, Lobanov V E,
and Torner L 2015 \emph{Opt. Lett.} \textbf{40}, 1045.


\bibitem{FR1} Papp S B, Pino J M, and Wieman C E 2008 \emph{Phys. Rev. Lett.}
\textbf{101}, 040402.

\bibitem{FR2} Zhang P, Naidon P, and Ueda M, 2009 \emph{Phys. Rev. Lett.}
\textbf{103}, 133202.


\bibitem{symbio1} P\'{e}rez-Garc\'{\i}a V M, and Belmonte Beitia J 2005
\emph{Phys. Rev. A} \textbf{72}, 033620.

\bibitem{symbio2} Adhikari S K, 2005 \emph{Phys. Lett. A} \textbf{346}, 179.

\bibitem{Townes} Chiao R Y, Garmire E, and Townes C H 1964 \emph{Phys. Rev.
Lett.} \textbf{15}, 1056.


\bibitem{Berge} Berg\'{e} L 1998 \emph{Phys. Rep.} \textbf{303}, 259.

\bibitem{Fibich} Fibich G \textit{The Nonlinear Schr\"{o}dinger Equation:
Singular Solutions and Optical Collapse} (Springer: Cham, 2015).

\bibitem{LHY} Lee T. D, Huang K, and Yang C N 1957 Phys. Rev. \textbf{106},
1135.


\bibitem{Petrov2015} Petrov D S 2015 \emph{Phys. Rev. Lett.} \textbf{115},
155302.

\bibitem{QDterm} Petrov D S, and Astrakharchik G E 2016 \emph{Phys. Rev.
Lett.} \textbf{117}, 100401.

\bibitem{Sadhan-nondip} Adhikari S K 2017 \emph{Phys. Rev. A} \textbf{95},
023606

\bibitem{Schutzhold2006} Sch\"{u}tzhold R, Uhlmann M, Xu Y, and Fischer U R
2006 \emph{Int. J. Mod. Phys. B} \textbf{20}, 3555.

\bibitem{Lima2011} Lima A R P and Pelster A 2011 \emph{Phys. Rev. A} \textbf{%
84}, 041604(R).

\bibitem{Saito2016} Saito H 2016 \emph{J. Phys. Soc. Jpn.} \textbf{85},
053001.

\bibitem{Bisset2016} Bisset R N, Wilson R M, Baillie D, and Blakie B P 2016
\emph{Phys. Rev. A} \textbf{94}, 033619.

\bibitem{Wachtler2016} W\"{a}chtler F, and Santos L 2016 \emph{Phys. Rev. A}
\textbf{93}, 061603(R).

\bibitem{Oldziejewski2016} O{\l }dziejewski and R, Jachymski K 2016 \emph{%
Phys. Rev. A} \textbf{94}, 063638.

\bibitem{Pastukhov2017} Pastukhov V 2017 \emph{Phys. Rev. A} \textbf{95},
023614.

\bibitem{Cinti2017} Cinti F, and Boninsegni M 2017 \emph{Phys. Rev. A}
\textbf{96}, 013627.

\bibitem{Sadhan-dip} Adhikari S K 2017 \emph{Laser Phys. Lett}. \textbf{14},
025501 (2017).

\bibitem{Chomaz2016} Chomaz L, Baier S, Petter D, Mark M J, W\"{a}chtler F,
Santos L, and Ferlaino F 2016 \emph{Phys. Rev. X} \textbf{6}, 041039.

\bibitem{Kadau2016} Kadau H, Schmitt M, Wenzel M, Wink C, Maier T,
Ferrier-Barbut I, and Pfau T 2016 \emph{Nature }(London) \textbf{530}, 194.

\bibitem{Ferrier2016} Ferrier-Barbut I, Kadau H, Schmitt M, Wenzel M, and
Pfau T 2016 \emph{Phys. Rev. Lett.} \textbf{116}, 215301.

\bibitem{Schmitt2016} Schmitt M, Wenzel M, B\"{o}ttcher F, Ferrier-Barbut I,
and Pfau T 2016 \emph{Nature} (London), \textbf{539}, 259.

\bibitem{Baillie2016} Baillie D, Wilson R M, Bisset R N, and Blakie P B 2016
\emph{Phys. Rev. A} \textbf{94}, 021602(R).

\bibitem{Edler2017} Edler D, Mishra C, W\"{a}chtler Nath F R, Sinha S, and
Santos L 2017 \emph{Phys. Rev. Lett} \textbf{119}, 050403.

\bibitem{Leticia} Cabrera C R , Tanzi L, Sanz J, Naylor B, Thomas P, Cheiney
P, and Tarruell L 2017 ePrint, arXiv:1708.07806.

\bibitem{Zhengwei2013} Zheng W, Yu Z, Cui X, and Zhai H 2013 \emph{J. Phys.
B: At. Mol. Opt. Phys.} \textbf{46} 134007.

\bibitem{Luca} Cappellaro A, Macr\'{\i} T, Bertacco G F, and Salasnich L,
2017 Sci. Rep. 7, 13358.

\bibitem{Agrawal} Agrawal G P \textit{Nonlinear Fiber Optics} (Academic
Press: San Diego, 1995).

\bibitem{Fermi} Astrakharchik G E, Boronat J, Casulleras J, and Giorgini S
2004 \emph{Phys. Rev. Lett.} \textbf{93}, 200404.

\bibitem{ITP1} Chiofalo L M, Succi S, and Tosi P M 2000 \emph{Phys. Rev. E}
\textbf{62}, 7438.

\bibitem{ITP2} Yang J, and Lakoba T I 2008 \emph{Stud. Appl. Math.} \textbf{%
120}, 265.

\bibitem{VaKo} Vakhitov M, and Kolokolov A 1973 \emph{Radiophys. Quantum
Electron.} \textbf{16}, 783.

\bibitem{SOM} Yang J, and Lakoba T I 2007 \emph{Stud. Appl. Math.} \textbf{%
118}, 153.


\bibitem{Sun2017} Sun F, Ye J, and Liu W M 2017 \emph{New J. Phys.} \textbf{%
19}, 063025.
\end{thebibliography}
\end{document}